\documentclass{WileyMSP-template}
\usepackage{amsmath}
\begin{document}

\pagestyle{fancy}

\title{Tuning the Coherent Propagation of Organic Exciton-Polaritons through the Cavity Q-factor}

\maketitle

% Author: Please give full first and last names for authors and include * after the name of all corresponding authors

\author{Ruth H. Tichauer$^{\dagger,}$*}
\author{Ilia Sokolovskii$^\dagger$}
\author{Gerrit Groenhof}

% Dedication

\dedication{}

% Affiliations: Please provide adacemic titles (Prof. or Dr.) for all authors where applicable, and include an institutional email address for all corresponding authors
\begin{affiliations}

Dr. R. H. Tichauer\\
Departamento de F\'{i}sica Te\'{o}rica de la Materia Condensada and Condensed Matter Physics Center (IFIMAC), Universidad Aut\'{o}noma de Madrid, Madrid, Spain.\\
Email Address: ruth.tichauer@uam.es

I. Sokolovskii, Prof. Dr. G. Groenhof\\
Nanoscience Center and Department of Chemistry, University of Jyv\"{a}skyl\"{a}, P.O. Box 35, 40014, Jyv\"{a}skyl\"{a}, Finland. \\
$^\dagger$equal contribution

\end{affiliations}

% Keywords: Please provide a minimum of three and a maximum of seven keywords, separated by commas


\begin{abstract}
Transport of excitons in organic materials can be enhanced through polariton formation when the interaction strength between these excitons and the confined light modes of an optical resonator  exceeds their decay rates. While the polariton lifetime is determined by the Q(uality)-factor of the optical resonator, the polariton group velocity is not. Instead, the latter is solely determined by the polariton dispersion. Yet, experiments suggest that the Q-factor also controls the polariton propagation velocity. To understand this observation, we performed molecular dynamics simulations of Rhodamine chromophores strongly coupled to Fabry-P\'{e}rot cavities with various Q-factors. Our results suggest that propagation in the aforementioned experiments is initially dominated by ballistic motion of upper polariton states at their group velocities, which leads to a rapid expansion of the wavepacket. Cavity decay in combination with non-adiabatic population transfer into dark states, rapidly depletes these bright states, causing the wavepacket to contract. However, because population transfer is reversible, propagation continues, but as a diffusion process, at lower velocity. By controlling the lifetime of bright states, the Q-factor determines the duration of the ballistic phase and the diffusion coefficient in the diffusive regime. Thus, polariton propagation in organic microcavities can be effectively tuned through the Q-factor.
\end{abstract}

\section{Introduction}

Achieving long-range energy transfer in organic media is a key requirement for enhancing the efficiency of opto-electronic devices, such as organic diodes or solar cells, in which energy transport is limited by the incoherent diffusion mechanism that governs the motion of Frenkel excitons through materials. Recent experiments suggest that strongly coupling such excitons to the confined, but \emph{delocalized}, modes of an optical resonator (called a cavity in what follows) can enhance transport through hybridization of the molecular excitons with the confined light modes into polaritons~\cite{Freixanet2000,Coles2014,Zhong2016,Zhong2017,Lerario2017,Myers2018,Rozenman2018,Zakharko2018,Georgiou2018,Xiang2020,Forrest2020,Pandya2021,Georgiou2021,Berghuis2022,Pandya2022}. 

Polaritons are coherent superpositions of molecular and cavity mode excitations that form when the interaction ($g$) between molecular excitons and cavity modes exceeds their decay rates ($\kappa_\text{mol}$ and $\gamma_\text{cav}$, respectively)~\cite{Torma2015,Forn-Diaz2019, Rider2022}. The vast majority of these light-matter hybrid states lack a strong contribution from the cavity mode excitations and are therefore ``dark". The few remaining states are bright and dispersive owing to their cavity mode contributions (Figure~\ref{fig:structure+dispersion}). Therefore, these bright states behave as quasi-particles with low effective mass and large group velocity~\cite{Agranovich2007}, which can be exploited for controlled and long-ranged {\it in-plane} energy transport~\cite{Litinskaya2008}. Indeed, {\it in-plane} polariton propagation has been observed in a variety of excitonic materials coupled to the confined light modes of Fabry-P\'{e}rot cavities~\cite{Rozenman2018, Pandya2022}, Bloch Surface Waves~\cite{Lerario2017,Forrest2020,Schwartz2022}, Surface Lattice Resonances~\cite{Berghuis2022}, and resonances arising from a dielectric constant mismatch between the excitonic medium and the surrounding environment~\cite{Pandya2021}.

While these observations are in line with theoretical predictions~\cite{Agranovich2007,Michetti2008, Litinskaya2008,Allard2022,Engelhardt2022}, the propagation velocity observed in these experiments, is significantly lower than the group velocities inferred from the polariton dispersion ($v_g=\partial\omega/\partial k_z$). In previous work~\cite{Sokolovskii2022} we used multi-scale molecular dynamics (MD) simulations to resolve this discrepancy, and showed that on long timescales ($>$ 100~fs) polariton propagation is a diffusion process. This diffusion is due to reversible population transfer between the stationary dark state manifold and the highly mobile bright polaritonic states, which renders the propagation speed much slower than the polariton group velocities. While we observed that cavity loss, caused by photon leakage through imperfect mirrors, reduces the distance over which polaritons propagate, we had not systematically investigated the effect of the cavity mode lifetime, $\tau_\text{cav}=\gamma_\text{cav}^{-1}$, which is related to the quality factor (Q-factor) via $\text{Q}=\omega_{\text{cav}}\tau_\text{cav}$.

The cavity mode lifetime, in combination with the molecular dephasing rate ($\kappa_\text{mol}$), determines how strong the light-matter interaction ($g$) needs to be for the molecule-cavity system to enter the strong coupling regime (for which various criteria are commonly employed~\cite{Rider2022}: (i) $g\ge \gamma_{\text{cav}}, \kappa_\text{mol}$; (ii) $g^2\ge (\gamma_{\text{cav}}-\kappa_\text{mol})^2/4$; (iii) $g^2\ge (\gamma_{\text{cav}}^2 + \kappa_\text{mol}^2)/2$; or (iv) $g\ge (\gamma_{\text{cav}} + \kappa_\text{mol})/2$). When strong coupling is achieved, the Q-factor only influences the lifetime of organic polaritons, but not the light-matter coupling strength~\cite{Tropf2017}. Therefore, the Q-factor neither affects the Rabi splitting between the lower (LP) and upper (UP) polariton branches ($\Omega^\text{Rabi}=2g\sqrt{N}$ with $N$ the number of molecules collectively coupled to the confined light modes, Figure~\ref{fig:structure+dispersion}), nor the group velocity of the bright polariton states. Yet, recent femtosecond transient absorption microscopy (fs-TAM) experiments by Pandya {\it{et al.}}~\cite{Pandya2022} on BODIPY-R dyes in Fabry-P\'{e}rot cavities with varying Q-factors, suggest a correlation between the observed polariton velocity and the cavity Q-factor.

%aim

To address this controversy and determine the effect of the cavity Q-factor on the propagation of organic polaritons, we 
performed atomistic MD simulations of Rhodamine chromophores strongly coupled to the confined light modes of one-dimensional (1D) uni-directional Fabry-P\'erot cavities~\cite{Michetti2005,Tichauer2021} with three different cavity mode lifetimes: $\tau_{\text{cav}} = $ 15, 30, and 60~fs. As before, the hydrated Rhodamines were modeled at the hybrid Quantum Mechanics / Molecular Mechanics (QM/MM) level~\cite{Warshel1976b,Boggio-Pasqua2012}. We calculated mean-field semi-classical MD trajectories of 512~molecules, including their solvent environment, strongly coupled to the 160~confined light modes of a red-detuned cavity (370~meV below the excitation energy of  Rhodamine, which is 4.18 eV at the CIS/3-21G//Amber03 level of theory employed here, see Computational Details and Supporting Information (SI) for details). Because in the fs-TAM measurements of Pandya {\it{et al.}}~\cite{Pandya2022} the 10~fs broad-band pump pulses populate mostly UP states, we modeled the initial excitation by preparing a Gaussian wavepacket of UP states centered at 
$\hbar\omega$= 4.41~eV with a bandwidth of $\sigma = $ 7.07~$\mu$m$^{-1}$~\cite{Agranovich2007}. The energy range of the states excited initially in this superposition is indicated by the magenta box in Figure~\ref{fig:structure+dispersion}{\bf{b}}.

\section{Results and Discussion}

% description of wavepacket, figure 2 and 3

In Figure~\ref{fig:wps_up} we show the time evolution of the probability density of the polaritonic wave function ($|\Psi(t)|^2$, Equation~\ref{eq:totalwf}), after instantaneous excitation of a Gaussian wavepacket of UP states in three Fabry-P\'{e}rot microcavities supporting cavity modes with 15, 30, and 60~fs lifetimes, and containing 512 Rhodamine molecules. Animations of the propagation of the total, molecular and photonic wavepackets are provided as SI.

In all cavities the wavepacket initially broadens due to the wide range of UP group velocities. Around 30~fs, however, the wavepacket splits into (i) a faster component with a short lifetime that depends on the Q-factor, and (ii) a slower component that is long-lived, but almost stationary. While the lifetime of the slower component is hardly affected by the cavity lifetime, its broadening is Q-factor dependent (Figure~\ref{fig:wps_up}a-f). The long lifetime of the slower part suggests that it is composed mostly of dark states that lack group velocity, and into which some population of the initially excited UP states has relaxed~\cite{Groenhof2019}. Nevertheless, due to thermally driven population transfer from these dark states back into propagating polaritons, the slower part still propagates. Because this transfer process is reversible and leads to transient occupation of polaritonic states over a wide range of $k_z$-vectors in both LP and UP branches, propagation occurs in a
diffusive manner~\cite{Sokolovskii2022}.

In contrast, the faster component of the wavepacket is mainly composed of the higher-energy UP states, which have high group velocity. Because the rate at which population transfers from these UP states into the dark state manifold is inversely proportional to the energy gap~\cite{Tichauer2022}, the main decay channel for these states is radiative emission through the imperfect cavity mirrors. Thus, the lifetime and hence propagation distance of the faster wavepacket component is Q-factor dependent, which is reflected by a faster rise of ground-state population when the cavity mode lifetime decreases (green dashed lines in Figure~\ref{fig:wps_up}{\bf{g}}-{\bf{i}}).

After the rapid initial expansion of the total wavepacket due to the population in the UP states (blue in Figure~\ref{fig:wps_up}{\bf{g}}-{\bf{i}}), transfer into the dark states (black), in combination with irreversible radiative decay from states with the highest group velocity, causes the wavepacket to contract. The extent of this contraction as well as the moment at which it takes place, depends on the cavity mode lifetime, as indicated by both the position, $\langle z\rangle$, and Mean Squared Displacement (MSD) of the wavepackets in Figure~\ref{fig:z_up}. 

Whereas during the expansion phase propagation is dominated by ballistic motion of fast UP states that reach longer distances for higher Q-factors (or equivalently, higher cavity mode lifetimes $\tau_{\text{cav}}$), as indicated by the maximum of the MSD ($\sim$~68, 23 and 7 $\mu$m$^2$), after contraction, propagation continues as diffusion, which is indicated by the linearity of the MSD at the end of the simulations (Figure~\ref{fig:z_up}). Diffusion emerges as a consequence of reversible population transfers between stationary dark states and mobile bright states at all $k_z$-vectors in both the UP and LP branches~\cite{Sokolovskii2022}. The turnover from ballistic propagation into diffusion is Q-factor dependent and occurs later when the cavity mode lifetime is higher (Figure~\ref{fig:z_up}). 

% analysis of the wavepacket in terms of MSD figure 4?

While simulations provide detailed mechanistic insights into polariton propagation, direct observation of such details is challenging experimentally, in particular because the multiple contributions to a single transient spectral signal of a molecule-cavity system cannot always be unambiguously disentangled~\cite{Renken2021}. In their fs-TAM experiments, Pandya {\it{et al.}}~\cite{Pandya2022} monitored the propagation of the wavepacket, $\Psi(z,t)$, by probing transient changes in cavity transmission at a wavelength that is sensitive to LP absorption. As explained in the SI, to mimic such pump-probe conditions in our simulations, we extracted position-dependent transient changes in the transmission from our trajectories as follows:
\begin{equation}
   \frac{\Delta T(z,t)}{T_0} = 
   \exp{\left(\varepsilon_\text{a}d|\Psi(z,t)|^2\right)}-1
   \label{eq:transmission}
\end{equation}
with $\Delta T(z,t) = T(z,t)-T_0$ the difference between $T(z,t)$, the transmission at position $z$ and time $t$ after excitation, and $T_0=T(z,0)$, the transmission before excitation. The variable $\varepsilon_\text{a}$ is the absorption coefficient and $d$ the path length. Because the value of $\varepsilon_\text{a}$ cannot be derived directly from MD simulations, we treated it together with $d$ as a single parameter. Here, we used $\varepsilon_\text{a}d$= 0.5, but, as we show in SI, varying this parameter does not change the results qualitatively. As was done in experiments~\cite{Rozenman2018,Pandya2022,Schwartz2022},
we characterize the propagation of the total wavepacket by the MSD of the transient signal, in our case of the transient transmission ($\Delta T/T_0$, Equation~\ref{eq:transmission}):
\begin{equation}
   \text{MSD}_T = \sum_i^N \left(z_i-z_0\right)^2\frac{\Delta T(z,t)}{T_0} = \sum_i^N\left(z_i-z_0\right)^2
   \left[\exp{\left(\varepsilon_\text{a}d|\Psi(z_i,t)|^2\right)}-1\right]
   \label{eq:MSD}
\end{equation}
with $z_0$ the expectation value of the position of the  wavepacket at the start of the simulation ($t$~=~0) and the sum is over the positions $z_i$ of the $N=512$ molecules. Full details of this analysis are provided in SI.

In Figure~\ref{fig:msd}{\bf{a}} we plot the MSD$_T$ of the transient differential transmission for our cavity systems. As in the experiments (Figure 2c in Pandya {\it{et al.}}~\cite{Pandya2022}), we observe that after a rapid initial increase, the MSD$_T$ of the signal decreases. Based on our simulations we attribute this observation to the fast expansion of the wavepacket followed by the contraction. Because two propagation regimes were observed in our simulations, we analysed these regimes separately. In contrast, Pandya {\it{et al.}} assumed a single ballistic phase, and extracted the velocity and duration of that phase from a global fit to the full MSD of the measured $\Delta T/T_0$ signal.

% ballistic stage

Because in the initial stages of the ballistic regime ($t<\tau_\text{cav}$) propagation is dominated by the population in UP states with well-defined dispersion, the propagation speed is independent of the Q-factor and determined solely by the UP group velocity (Figure~\ref{fig:structure+dispersion}{\bf{c}}) in all cavities (Figure S2{\bf{b}} in SI). However, the duration of this ballistic regime, $\tau_{\text{bal}}$, extracted from the  $\Delta T/T_0$ signal by fitting the same function as Pandya {\it{et al.}} to the initial rise of the MSD$_T$ (SI), depends on the cavity lifetime, and lasts longer if the cavity Q-factor is higher, as shown in Figure~\ref{fig:msd}{\bf{b}}. Therefore, as in the MSD plots of the total wavepacket in Figure~\ref{fig:z_up}, the MSD$_T$ of the $\Delta T/T_0$ signal also reaches the highest value in the cavity with highest Q-factor (or equivalently, the longest cavity mode lifetime $\tau_{\text{cav}}$), in line with the fs-TAM measurements.

% diffusion stage

Whereas in their model Pandya {\it{et al.}} consider only ballistic propagation on a sub-ps timescale, our simulations suggest that also diffusion contributes to propagation on those timescales, when solely the slower part of the wavepacket remains. Therefore, to characterize also this regime, we calculated the diffusion coefficient by fitting the linear regime of the MSD (SI). However, because in the MSD$_T$ of the transient transmission (Figure~\ref{fig:msd}{\bf{a}}), the linear regime is difficult to discern, we performed the linear fit to the MSD associated with the slower component of the wavepacket at the end of the trajectories (Figure~S5 in SI). In Figure~\ref{fig:msd}{\bf{c}} we plot the diffusion coefficients as a function of cavity mode lifetime. Since the Q-factor determines the lifetime of the population in the propagating bright states, the diffusion constant increases with the cavity lifetime. 

Because in our simulations we cannot couple as many molecules to the cavity as in experiment ({\it{i.e.}}, 10$^5$-10$^8$ molecules~\cite{Houdre1996,Eizner2019,Martinez2019}), we overestimate the diffusion coefficient. As we could show previously~\cite{Tichauer2022}, the rate of population transfer from dark to bright states is inversely proportional to $N$, whereas the rate in the opposite direction is independent of $N$. Therefore, the population in the bright propagating states is overestimated when only 512~molecules are coupled to the cavity, leading to a faster diffusion.  The overestimation of the diffusion coefficient thus leads to a much more pronounced increase of the wavepacket MSD than in experiment, where the total population residing in the propagating states is significantly lower~\cite{Schwartz2022}, and  diffusion would be hardly observable on sub-ps timescales. Nevertheless, despite these quantitative  differences, our simulations provide a qualitative picture that is in line with experimental observations~\cite{Pandya2022}.

% wrapping up 

The results of our simulations suggest that the cavity lifetime controls both the duration and length of the initial ballistic phase (Figure~\ref{fig:msd}{\bf{b}}) as well as the diffusion constant in the diffusive regime (Figure~\ref{fig:msd}{\bf{c}}). Thus, without affecting polariton group velocity, the cavity Q-factor provides an effective means to tune energy transport in the strong coupling regime.

\section{Conclusion}

To summarize, we have investigated the effect of the cavity Q-factor on polariton propagation by means of atomistic MD simulations. In line with experiments, we find that the Q-factor determines the propagation velocity and distance of organic polaritons via their lifetimes without affecting group velocities. Our findings therefore resolve the unexpected correlation between Q-factor and propagation velocity reported by Pandya {\it et al.}~\cite{Pandya2022}. Our results furthermore underscore that to understand the mechanism of polariton propagation and interpret experiments, it is necessary to include: ({\it i}) atomic details for the material; ({\it ii})  multiple modes for cavity dispersion;  ({\it iii}) cavity decay; and ({\it iv}) sufficiently many molecules to have dark states providing an exciton reservoir. In particular, treating the molecular degrees of freedom of many molecules is essential for observing wavepacket contraction that is caused by cavity loss in combination with reversible non-adiabatic population transfer between propagating bright states and the stationary long-lived dark state manifold. Our work suggests that an {\it ab initio} description of molecules in multi-mode cavities could pave the way to systematically design or optimize polariton-based devices for enhanced energy transport.

\section{Computational Details}

We performed mean-field semi-classical~\cite{Ehrenfest1927} MD simulations of 512 Rhodamine chromophores with their solvent environment, strongly coupled to 1D Fabry-P\'erot cavities with different radiative lifetimes: $\tau_{\text{cav}} = $ 15~fs, 30~fs, and 60~fs. To model the interactions between the molecules and the confined light modes of the cavity, we used a Tavis-Cummings Hamiltonian, in which the molecular degrees of freedom are included~\cite{Luk2017,Tichauer2021}. A brief description of our multi-scale cavity MD approach is provided as SI.

In our simulations the Rhodamine molecules were modelled at the QM/MM level, with the QM region containing the fused ring system of the molecule (Figure~S2 in SI). The ground-state electronic structure of the QM subsystem was described at the restricted Hartree-Fock (HF) method in combination with the 3-21G basis set~\cite{Ditchfield1971}, while the excited-state electronic structure was modeled with Configuration Interaction, truncated at single electron excitations (CIS/3-21G). The MM region, which contains the rest of the chromophore as well as the solvent (3684 water molecules), was modeled with the Amber03 force field~\cite{Duan2003} in combination with the TIP3P water model~\cite{Jorgensen1983}. At this level of QM/MM theory, the excitation energy of the Rhodamine molecules is 4.18~eV~\cite{Luk2017}. In previous work, we showed that despite the overestimation of the vertical excitation energy, the topology of the potential energy surfaces is not very sensitive to the level of theory for Rhodamine~\cite{Groenhof2019}.

\begin{figure}
  \includegraphics[width=12cm]{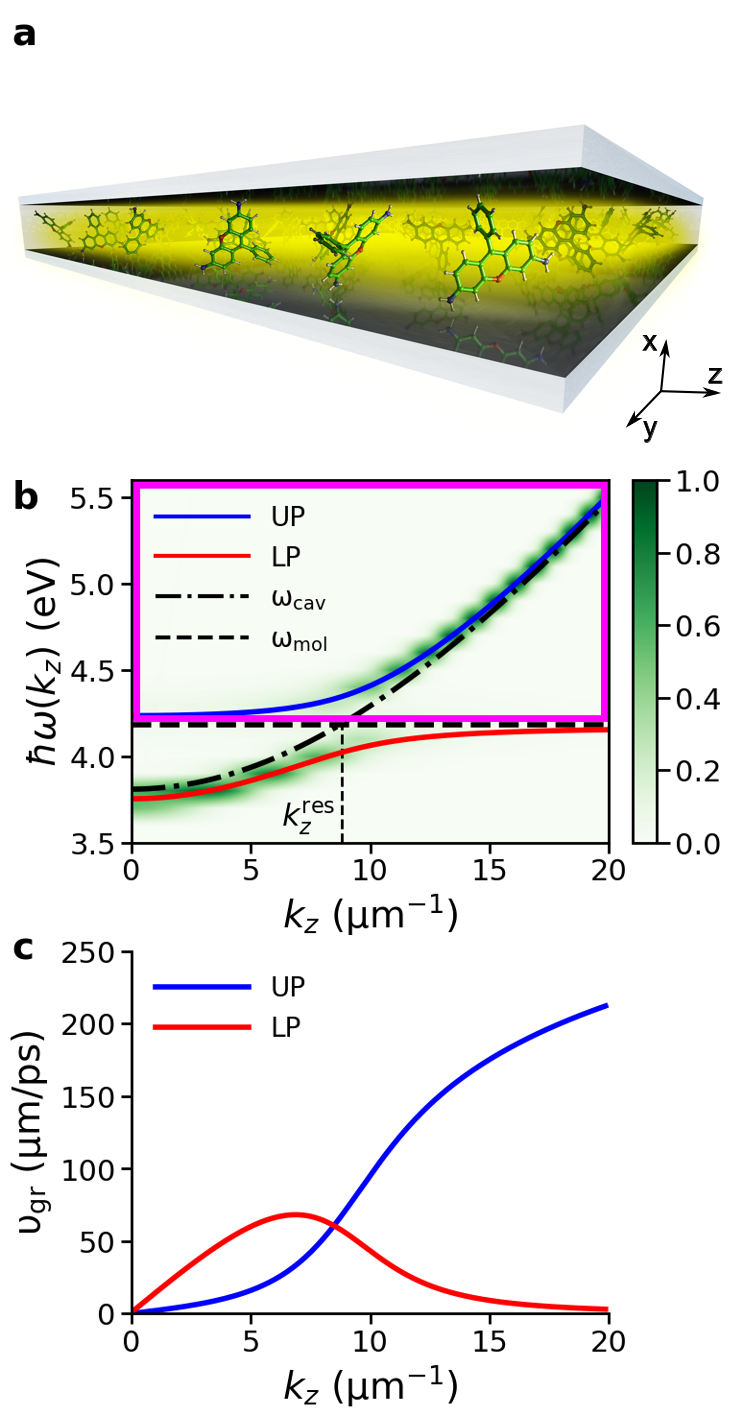}
  \caption{Panel {\bf{a}}: Schematic illustration of an optical Fabry-P\'erot micro-cavity filled with Rhodamine chromophores (not to scale). Panel {\bf b}: Normalised angle-resolved absorption spectrum of the cavity, showing Rabi splitting between the lower polariton (LP, red line) and the upper polariton (UP, blue line) branches. The cavity dispersion and excitation energy of the molecules (4.18~eV at the CIS/3-21G//Amber03 level of theory) are plotted by point-dashed and dashed lines, respectively. The purple frame encloses the range of polaritonic states excited instantaneously by the broad-band pump pulse. Panel {\bf c}: Group velocity of the LP (red) and UP (blue), defined as $\partial\omega(k_z)/\partial k_z$.}\label{fig:structure+dispersion}
\end{figure}

\begin{figure}
  \includegraphics[width=\linewidth]{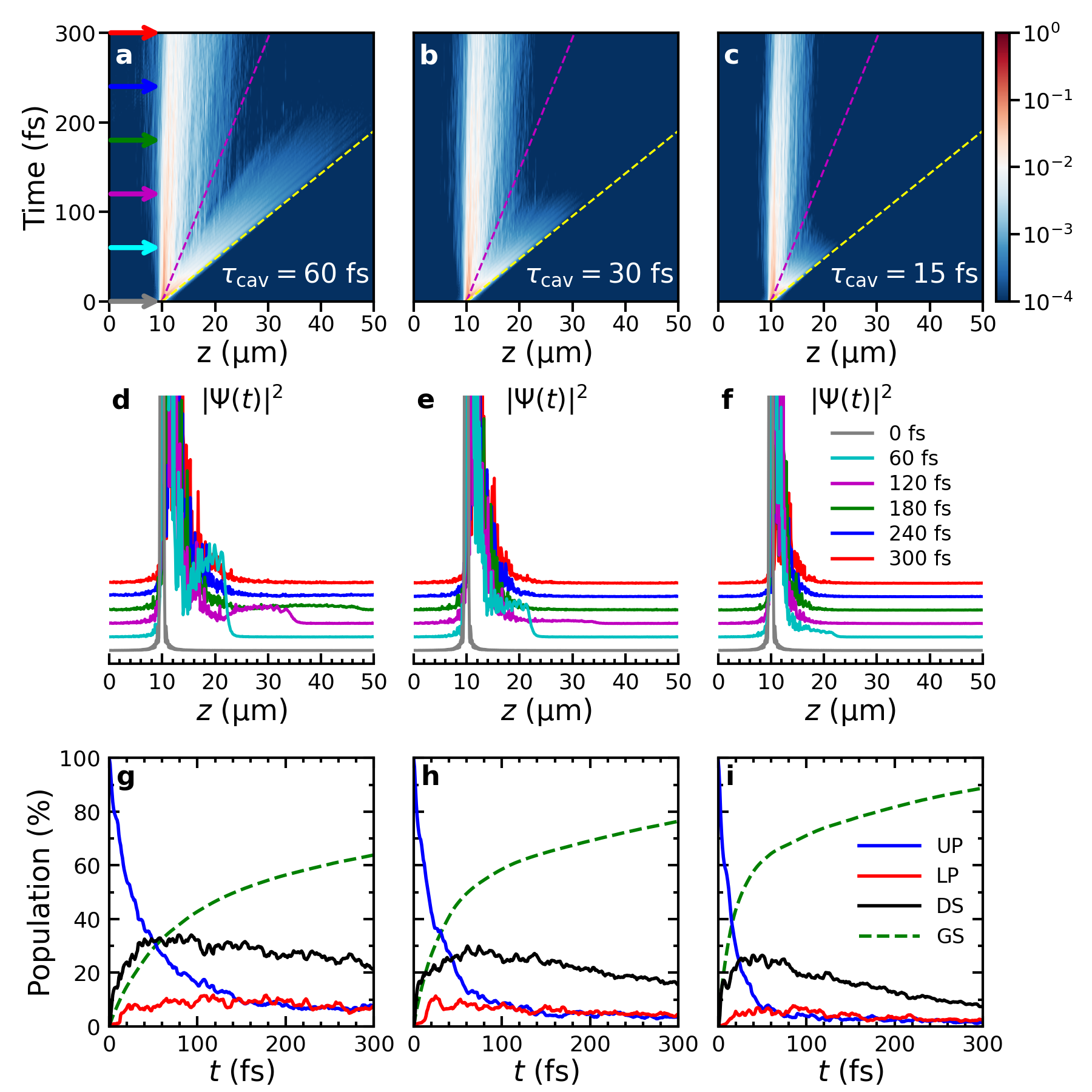}
  \caption{Polariton propagation after resonantly exciting a wavepacket of states in the UP branch centered at z=10~$\mu$m. Panels {\bf{a}}, {\bf{b}} and {\bf{c}}: probability density of the total wave function, $|\Psi(t)|^2$, as a function of distance (horizontal-axis) and time (vertical-axis) in cavities with different Q-factors ({\it i.e.}, $\tau_{\text{cav}} = 60, 30$ and $15~\text{fs}$, respectively). Colored arrows in panel {\bf{a}} correspond to the time points of the 1D projection in panels {\bf{d}}, {\bf{e}} and {\bf{f}}. The dashed purple and yellow lines indicate propagation at the maximum group velocity of the LP (68~$\mu$m ps$^{-1}$) and UP (212~$\mu$mps$^{-1}$) branches, respectively. Panels {\bf{d}}, {\bf{e}} and {\bf{f}}: probability density of the total polariton wave function, $|\Psi(t)|^2$, at different time points as a function of distance. Panels {\bf{g}}, {\bf{h}}, and {\bf{i}}:  
  populations of the UP (blue), LP (red), and dark (DS, black) states, as well as of the ground state (GS, green dashed line) as functions of time.}
 \label{fig:wps_up}
\end{figure}

\begin{figure}
  \includegraphics[width=\linewidth]{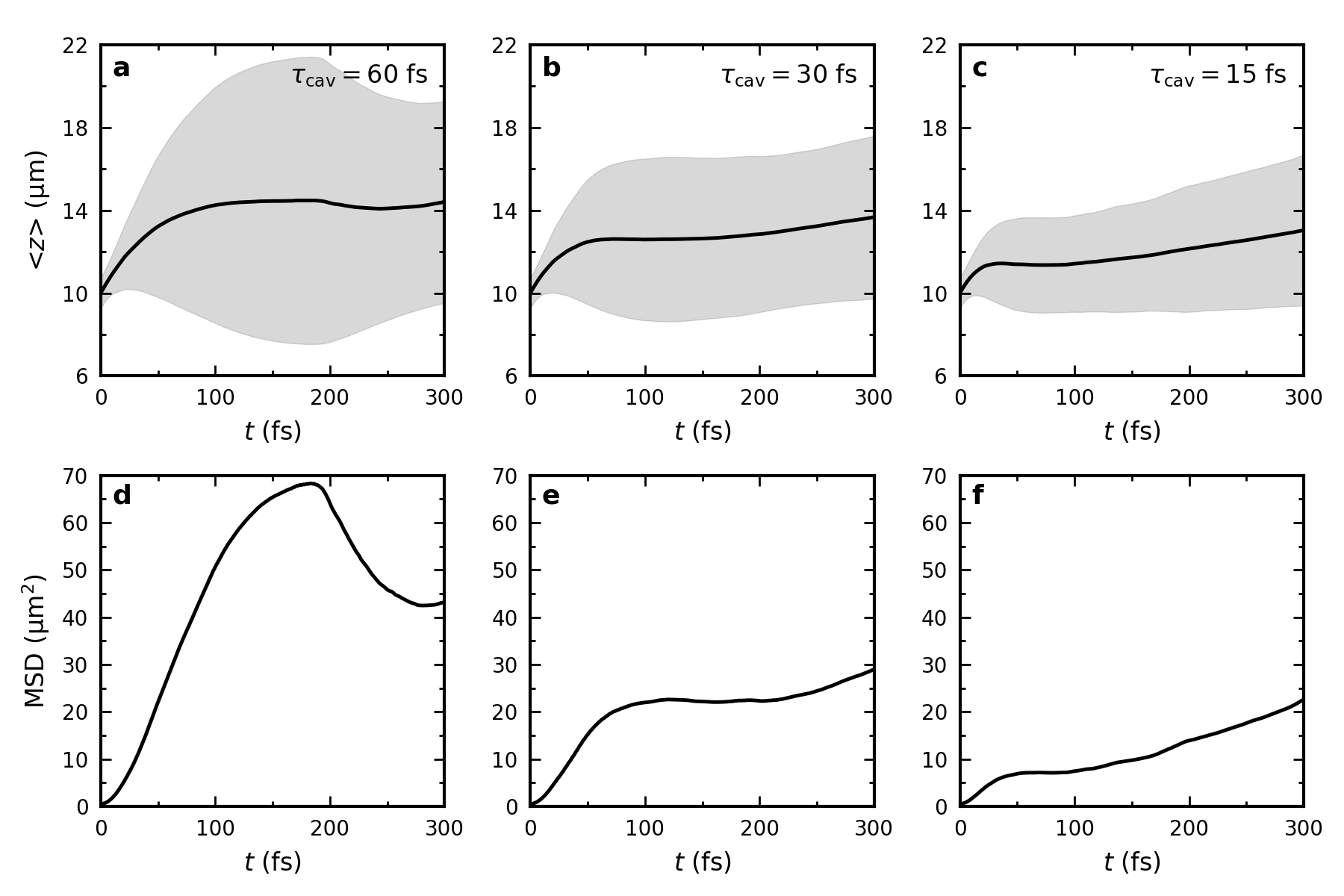}
  \caption{Top panels: Expectation value of the position of the total-time dependent wavefunction $\langle z\rangle=\langle\Psi(t)|\hat{z}(t)|\Psi(t)\rangle/\langle\Psi(t)|\Psi(t)\rangle$ after on-resonant excitation of UP states in cavities with different Q-factors ({\it i.e.}, $\tau_{\text{cav}} =$~60 (left), 30~(middle) and 15~fs~(right). The black lines represent $\langle z\rangle$ while the shaded areas indicate the root mean squared deviation (RMSD, {\it{i.e.}}, $\sqrt{\langle (z(t)-\langle z(t)\rangle)^2\rangle}$). Bottom panels: Mean squared displacement (MSD, {\it{i.e.}}, $\langle\Psi(t)|\left(\hat{z}(t)-\hat{z}(0)\right)^2|\Psi(t)\rangle/\langle\Psi(t)|\Psi(t)\rangle$) in the same cavities. 
  }
  \label{fig:z_up}
\end{figure}

\begin{figure}
  \includegraphics[width=\linewidth]{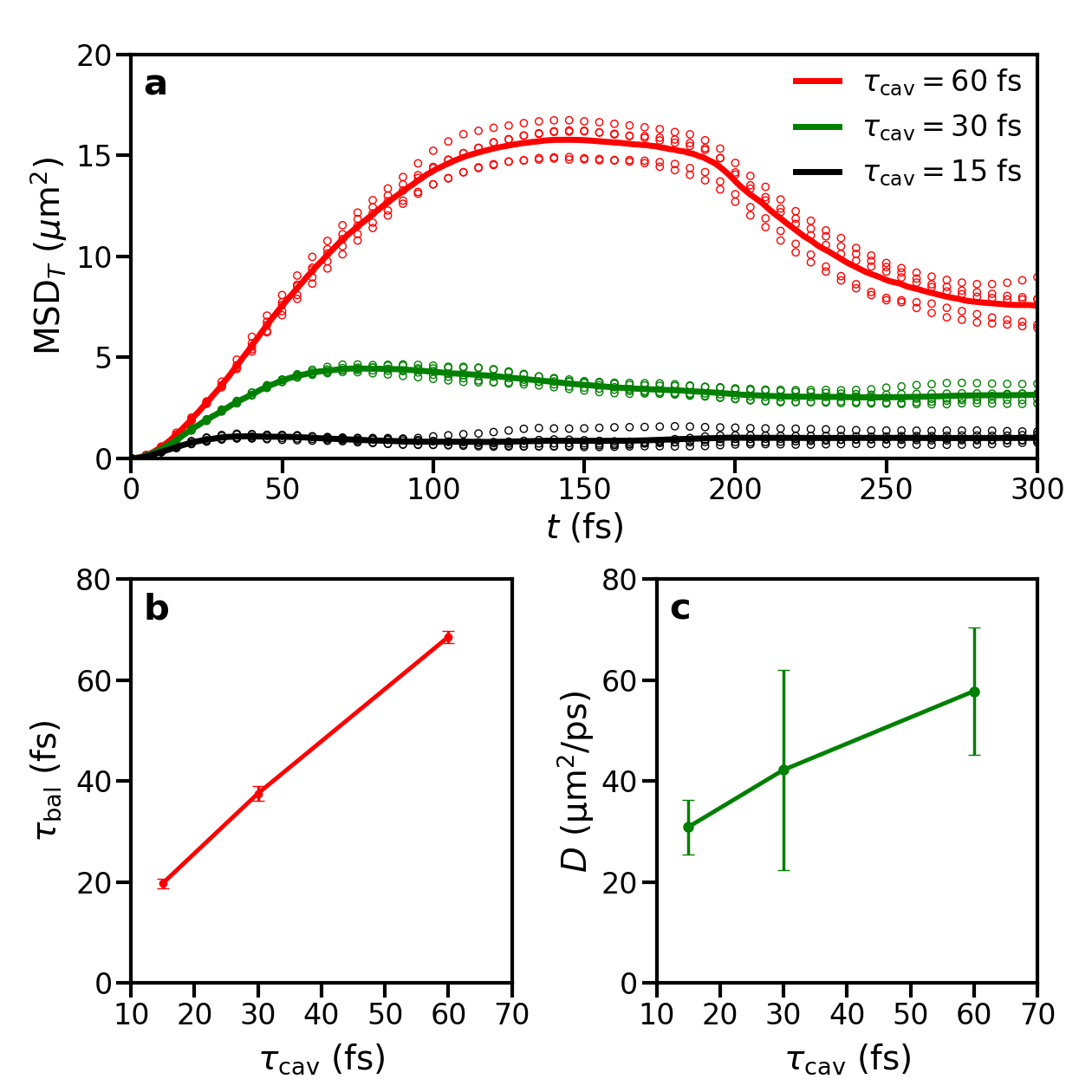}
 \caption{Panel {\bf{a}}: Mean squared displacement of the transmission signal ($\text{MSD}_T$) for different cavities mode lifetimes: $\tau_{\text{cav}} = 60$ (red), 30 (green) and $15~\text{fs}$ (black). Circles represent data points for individual runs, while the curves show the averages over all trajectories (five for each $\tau_{\text{cav}}$). Panel {\bf{b}}: the duration of the ballistic phase as a function of cavity mode lifetime. Panel {\bf{c}}: the diffusion coefficient in the diffusion regime as a function of cavity mode lifetime.
}

\label{fig:msd}
\end{figure}

The uni-directional 1D cavity with a length of $L_z = $~50~$\mu$m, with $z$ indicating the in-plane direction ($L_x = 163$~nm is the distance between the mirrors and $x$ thus indicates the out-of-plane direction, see Figure S1 in the SI), was red-detuned by 370~meV with respect to the molecular excitation energy (4.18~eV at the CIS/3-21G//Amber03 level of theory, dashed line in Figure~\ref{fig:structure+dispersion}{\bf b}), such that at wave vector $k_z = 0$, the cavity resonance is  $\hbar\omega_0 =$~3.81~eV. The cavity dispersion, $\omega_\text{cav}(k_{z}) = \sqrt{\omega_0^2+c^2k_{z}^2/n^2}$, was modelled with 160 modes ($0\le p\le 159$ for $k_{z}=2\pi p/L_z$), with $c$ the speed of light and $n$ the refractive index. Here, we used $n=1$. See SI for further details on the cavity model.

The Rhodamine molecules were placed with equal inter-molecular distances on the $z$-axis of the cavity. To maximize the collective light-matter coupling strength, the transition dipole moments of the Rhodamine molecules were aligned to the vacuum field at the start of the simulation. 
The same starting coordinates were used for all Rhodamines, but different initial velocities were selected randomly from a Maxwell-Boltzmann distribution at 300~K. 

With a cavity vacuum field strength of 0.36~MVcm$^{-1}$ (0.0000707~au), the Rabi splitting, defined as the energy difference between the bright lower (LP) and upper polariton (UP) branches at the wave-vector $k^{\text{res}}_z$ where the cavity dispersion matches the molecular excitation energy (Figure~\ref{fig:structure+dispersion}{\bf b}), is $\sim$~325~meV for all cavities ($\tau_{\text{cav}} = $ 15~fs, 30~fs, and 60~fs). While the choice for a 1D cavity model with only positive $k_z$ vectors was motivated by the necessity to keep our simulations computationally tractable, it precludes the observation of elastic scattering events that would change the direction ({\it{i.e.}}, in-plane momentum, $\hbar{\bf{k}}$) of propagation. Furthermore, with only positive $k_z$ vectors, polariton motion is restricted to the $+z$ direction, but we could show previously~\cite{Sokolovskii2022} that this assumption does not affect the mechanism of the propagation process.

Ehrenfest MD trajectories were computed by numerically integrating Newton's equations of motion using a leap-frog algorithm with a 0.1~fs timestep. The multi-mode Tavis-Cummings Hamiltonian (See SI) was diagonalized at each time-step to obtain the $N+n_\text{mode}$ (adiabatic) polaritonic eigenstates $|\psi^m\rangle$ and energies $E^m$.
The \emph{total} polaritonic wavefunction $|\Psi(t)\rangle$ was coherently propagated along with the classical degrees of freedom of the $N$ molecules as a time-dependent superposition of the $N+n_\text{mode}$ time-independent adiabatic polaritonic states:
\begin{equation}
|\Psi(t)\rangle=\sum_m^{N+n_{\text{mode}}}c_m(t)|\psi ^m\rangle\label{eq:totalwf}
\end{equation}
where $c_m(t)$ are the time-dependent expansion coefficients of the time-independent polaritonic eigenstates $|\psi^m\rangle$ (SI). A unitary propagator in the \emph{local} diabatic basis was used to integrate these coefficients~\cite{Granucci2001}, while the nuclear degrees of freedom of the $N$ molecules evolve on the mean-field potential energy surface. Results reported in this work were obtained as averages over five trajectories for each cavity lifetime. For all  simulations we used Gromacs~4.5.3~\cite{Hess2008}, in which the multi-mode Tavis-Cummings QM/MM model was implemented~\cite{Tichauer2021}, in combination with Gaussian16~\cite{g16}. Further details of the simulations are provided in the SI. 

\medskip
\textbf{Supporting Information} \par 
Supporting Information (SI) is available.

% Acknowledgements
\medskip
\textbf{Acknowledgements} \par %delete if not applicable))
R.H.T and I.S. contributed equally to this work. The authors thank Dmitry Morozov for his help during the project. I.S. and G.G. thank Pavel Buslaev for valuable discussions. We also thank CSC-IT center for scientific computing in Espoo, Finland, for very generous computational resources, and for assistance in running the simulations. This work was supported by the Academy of Finland (Grant 323996), the European Research Council (Grant No. ERC-2016-StG-714870 to Johannes Feist), and by the Spanish Ministry for Science, Innovation, Universities-Agencia Estatal de Investigaci\'{o}n (AEI) through Grants (PID2021-125894NB-I00 and CEX2018-000805-M (through the Mar\'{i}a de Maeztu program for Units of Excellence in Research and Development).
% References
\medskip

% Use the following code if you wish to generate your bibliography with BibTeX;
% replace the string "MSP-template" below with the name(s) of
% the BibTeX data base(s) you want to use.
% The resulting bibliography-output (the content of the .bbl file)
% must be pasted back into this file before submission.
% Please also include your BibTeX data base file(s) in your submission
% so that we can re-run BibTeX if necessary.
%

\bibliographystyle{MSP}
\bibliography{main}

\providecommand{\latin}[1]{#1}
\makeatletter
\providecommand{\doi}
  {\begingroup\let\do\@makeother\dospecials
  \catcode`\{=1 \catcode`\}=2 \doi@aux}
\providecommand{\doi@aux}[1]{\endgroup\texttt{#1}}
\makeatother
\providecommand*\mcitethebibliography{\thebibliography}
\csname @ifundefined\endcsname{endmcitethebibliography}
  {\let\endmcitethebibliography\endthebibliography}{}
\begin{mcitethebibliography}{23}
\providecommand*\natexlab[1]{#1}
\providecommand*\mciteSetBstSublistMode[1]{}
\providecommand*\mciteSetBstMaxWidthForm[2]{}
\providecommand*\mciteBstWouldAddEndPuncttrue
  {\def\EndOfBibitem{\unskip.}}
\providecommand*\mciteBstWouldAddEndPunctfalse
  {\let\EndOfBibitem\relax}
\providecommand*\mciteSetBstMidEndSepPunct[3]{}
\providecommand*\mciteSetBstSublistLabelBeginEnd[3]{}
\providecommand*\EndOfBibitem{}
\mciteSetBstSublistMode{f}
\mciteSetBstMaxWidthForm{subitem}{(\alph{mcitesubitemcount})}
\mciteSetBstSublistLabelBeginEnd
  {\mcitemaxwidthsubitemform\space}
  {\relax}
  {\relax}

\bibitem[Jaynes and Cummings(1963)Jaynes, and Cummings]{Jaynes1963}
Jaynes,~E.~T.; Cummings,~F.~W. Comparison of quantum and semiclassical
  radiation theories with to the beam maser. \emph{Proc. IEEE} \textbf{1963},
  \emph{51}, 89--109\relax
\mciteBstWouldAddEndPuncttrue
\mciteSetBstMidEndSepPunct{\mcitedefaultmidpunct}
{\mcitedefaultendpunct}{\mcitedefaultseppunct}\relax
\EndOfBibitem
\bibitem[Tavis and Cummings(1969)Tavis, and Cummings]{Tavis1969}
Tavis,~M.; Cummings,~F.~W. Approximate solutions for an N-molecule
  radiation-field Hamiltonian. \emph{Phys. Rev.} \textbf{1969}, \emph{188},
  692--695\relax
\mciteBstWouldAddEndPuncttrue
\mciteSetBstMidEndSepPunct{\mcitedefaultmidpunct}
{\mcitedefaultendpunct}{\mcitedefaultseppunct}\relax
\EndOfBibitem
\bibitem[Luk \latin{et~al.}(2017)Luk, Feist, Toppari, and Groenhof]{Luk2017}
Luk,~H.-L.; Feist,~J.; Toppari,~J.~J.; Groenhof,~G. Multiscale Molecular
  Dynamics Simulations of Polaritonic Chemistry. \emph{J. Chem. Theory Comput.}
  \textbf{2017}, \emph{13}, 4324--4335\relax
\mciteBstWouldAddEndPuncttrue
\mciteSetBstMidEndSepPunct{\mcitedefaultmidpunct}
{\mcitedefaultendpunct}{\mcitedefaultseppunct}\relax
\EndOfBibitem
\bibitem[Tichauer \latin{et~al.}(2021)Tichauer, Feist, and
  Groenhof]{Tichauer2021}
Tichauer,~R.; Feist,~J.; Groenhof,~G. Multi-scale Dynamics Simulations of
  Molecular Polaritons: the Effect of Multiple Cavity Modes on Polariton
  Relaxation. \emph{J. Chem. Phys} \textbf{2021}, \emph{154}, 104112\relax
\mciteBstWouldAddEndPuncttrue
\mciteSetBstMidEndSepPunct{\mcitedefaultmidpunct}
{\mcitedefaultendpunct}{\mcitedefaultseppunct}\relax
\EndOfBibitem
\bibitem[Warshel and Levitt(1976)Warshel, and Levitt]{Warshel1976b}
Warshel,~A.; Levitt,~M. Theoretical studies of enzymatic reactions: Dielectric,
  electrostatic and steric stabilization of carbonium ion in the reaction of
  lysozyme. \emph{J. Mol. Biol.} \textbf{1976}, \emph{103}, 227--249\relax
\mciteBstWouldAddEndPuncttrue
\mciteSetBstMidEndSepPunct{\mcitedefaultmidpunct}
{\mcitedefaultendpunct}{\mcitedefaultseppunct}\relax
\EndOfBibitem
\bibitem[Boggio-Pasqua \latin{et~al.}(2012)Boggio-Pasqua, Burmeister, Robb, and
  Groenhof]{Boggio-Pasqua2012}
Boggio-Pasqua,~M.; Burmeister,~C.~F.; Robb,~M.~A.; Groenhof,~G. Photochemical
  reactions in biological systems: probing the effect of the environment by
  means of hybrid quantum chemistry/molecular mechanics simulations.
  \emph{Phys. Chem. Chem. Phys.} \textbf{2012}, \emph{14}, 7912--7928\relax
\mciteBstWouldAddEndPuncttrue
\mciteSetBstMidEndSepPunct{\mcitedefaultmidpunct}
{\mcitedefaultendpunct}{\mcitedefaultseppunct}\relax
\EndOfBibitem
\bibitem[Michetti and Rocca(2005)Michetti, and Rocca]{Michetti2005}
Michetti,~P.; Rocca,~G. C.~L. Polariton states in disordered organic
  microcavities. \emph{Phys. Rev. B.} \textbf{2005}, \emph{71}, 115320\relax
\mciteBstWouldAddEndPuncttrue
\mciteSetBstMidEndSepPunct{\mcitedefaultmidpunct}
{\mcitedefaultendpunct}{\mcitedefaultseppunct}\relax
\EndOfBibitem
\bibitem[Agranovich and Gartstein(2007)Agranovich, and
  Gartstein]{Agranovich2007}
Agranovich,~V.; Gartstein,~Y. Nature and Dynamics of Low-Energy Exciton
  Polaritons in Semiconductor Microcavities. \emph{Phys. Rev. B} \textbf{2007},
  \emph{75}, 075302\relax
\mciteBstWouldAddEndPuncttrue
\mciteSetBstMidEndSepPunct{\mcitedefaultmidpunct}
{\mcitedefaultendpunct}{\mcitedefaultseppunct}\relax
\EndOfBibitem
\bibitem[Granucci \latin{et~al.}(2001)Granucci, Persico, and
  Toniolo]{Granucci2001}
Granucci,~G.; Persico,~M.; Toniolo,~A. Direct semiclassical simulation of
  photochemical processes with semiempirical wave functions. \emph{J. Chem.
  Phys.} \textbf{2001}, \emph{114}, 10608--10615\relax
\mciteBstWouldAddEndPuncttrue
\mciteSetBstMidEndSepPunct{\mcitedefaultmidpunct}
{\mcitedefaultendpunct}{\mcitedefaultseppunct}\relax
\EndOfBibitem
\bibitem[Duan \latin{et~al.}({2003})Duan, Wu, Chowdhury, Lee, Xiong, Zhang,
  Yang, Cieplak, Luo, Lee, Caldwell, Wang, and Kollman]{Duan2003}
Duan,~Y.; Wu,~C.; Chowdhury,~S.; Lee,~M.~C.; Xiong,~G.~M.; Zhang,~W.; Yang,~R.;
  Cieplak,~P.; Luo,~R.; Lee,~T.; Caldwell,~J.; Wang,~J.~M.; Kollman,~P. {A
  point-charge force field for molecular mechanics simulations of proteins
  based on condensed-phase quantum mechanical calculations}. \emph{J. Comput.
  Chem.} \textbf{{2003}}, \emph{{24}}, {1999--2012}\relax
\mciteBstWouldAddEndPuncttrue
\mciteSetBstMidEndSepPunct{\mcitedefaultmidpunct}
{\mcitedefaultendpunct}{\mcitedefaultseppunct}\relax
\EndOfBibitem
\bibitem[Jorgensen \latin{et~al.}(1983)Jorgensen, Chandrasekhar, Madura, Impey,
  and Klein]{Jorgensen1983}
Jorgensen,~W.~L.; Chandrasekhar,~J.; Madura,~J.~D.; Impey,~R.~W.; Klein,~M.~L.
  Comparison of simple potential functions for simulatin liquid water. \emph{J.
  Chem. Phys.} \textbf{1983}, \emph{79}, 926--935\relax
\mciteBstWouldAddEndPuncttrue
\mciteSetBstMidEndSepPunct{\mcitedefaultmidpunct}
{\mcitedefaultendpunct}{\mcitedefaultseppunct}\relax
\EndOfBibitem
\bibitem[Berendsen \latin{et~al.}(1984)Berendsen, Postma, van Gunsteren, la,
  and Haak]{Berendsen1984}
Berendsen,~H.; Postma,~J.; van Gunsteren,~W.; la,~A.~D.; Haak,~J. Molecular
  dynamics with coupling to an external bath. \emph{J. Chem. Phys.}
  \textbf{1984}, \emph{81}, 3684--3690\relax
\mciteBstWouldAddEndPuncttrue
\mciteSetBstMidEndSepPunct{\mcitedefaultmidpunct}
{\mcitedefaultendpunct}{\mcitedefaultseppunct}\relax
\EndOfBibitem
\bibitem[Hess \latin{et~al.}(1997)Hess, Bekker, Berendsen, and
  Fraaije]{Hess1997}
Hess,~B.; Bekker,~H.; Berendsen,~H. J.~C.; Fraaije,~J. G. E.~M. {LINCS: A
  linear constraint solver for molecular simulations}. \emph{J. Comput. Chem.}
  \textbf{1997}, \emph{18}, 1463--1472\relax
\mciteBstWouldAddEndPuncttrue
\mciteSetBstMidEndSepPunct{\mcitedefaultmidpunct}
{\mcitedefaultendpunct}{\mcitedefaultseppunct}\relax
\EndOfBibitem
\bibitem[Miyamoto and Kollman(1992)Miyamoto, and Kollman]{Miyamoto1992}
Miyamoto,~S.; Kollman,~P.~A. {SETTLE}: An analytical version of the {SHAKE} and
  {RATTLE} algorithms for rigid water molecules. \emph{J. Comp. Chem.}
  \textbf{1992}, \emph{18}, 1463--1472\relax
\mciteBstWouldAddEndPuncttrue
\mciteSetBstMidEndSepPunct{\mcitedefaultmidpunct}
{\mcitedefaultendpunct}{\mcitedefaultseppunct}\relax
\EndOfBibitem
\bibitem[Essmann \latin{et~al.}(1995)Essmann, Perera, Berkowitz, Darden, Lee,
  and Pedersen]{Essmann1995}
Essmann,~U.; Perera,~L.; Berkowitz,~M.~L.; Darden,~T.; Lee,~H.; Pedersen,~L.~G.
  A smooth particle mesh {E}wald potential. \emph{J. Chem. Phys} \textbf{1995},
  \emph{103}, 8577--8592\relax
\mciteBstWouldAddEndPuncttrue
\mciteSetBstMidEndSepPunct{\mcitedefaultmidpunct}
{\mcitedefaultendpunct}{\mcitedefaultseppunct}\relax
\EndOfBibitem
\bibitem[Groenhof \latin{et~al.}(2019)Groenhof, Climent, Feist, Morozov, and
  Toppari]{Groenhof2019}
Groenhof,~G.; Climent,~C.; Feist,~J.; Morozov,~D.; Toppari,~J.~J. Tracking
  Polariton Relaxation with Multiscale Molecular Dynamics Simulations. \emph{J.
  Chem. Phys. Lett.} \textbf{2019}, \emph{10}, 5476--5483\relax
\mciteBstWouldAddEndPuncttrue
\mciteSetBstMidEndSepPunct{\mcitedefaultmidpunct}
{\mcitedefaultendpunct}{\mcitedefaultseppunct}\relax
\EndOfBibitem
\bibitem[Groenhof \latin{et~al.}(2004)Groenhof, Bouxin-Cademartory, Hess,
  de~Visser, Berendsen, Olivucci, and A.~E.~Mark]{Groenhof2004}
Groenhof,~G.; Bouxin-Cademartory,~M.; Hess,~B.; de~Visser,~S.~P.; Berendsen,~H.
  J.~C.; Olivucci,~M.; A.~E.~Mark,~M. A.~R. Photoactivation of the photoactive
  yellow protein: Why photon absorption triggers a trans-to-cis isomerization
  of the chromophore in the protein. \emph{J. Am. Chem. Soc} \textbf{2004},
  \emph{124}, 4228--4232\relax
\mciteBstWouldAddEndPuncttrue
\mciteSetBstMidEndSepPunct{\mcitedefaultmidpunct}
{\mcitedefaultendpunct}{\mcitedefaultseppunct}\relax
\EndOfBibitem
\bibitem[Hess \latin{et~al.}(2008)Hess, Kutzner, van~der Spoel, and
  Lindahl]{Hess2008}
Hess,~B.; Kutzner,~C.; van~der Spoel,~D.; Lindahl,~E. GROMACS 4: Algorithms for
  Highly Efficient, Load-Balanced, and Scalable Molecular Simulation. \emph{J.
  Chem. Theory Comput.} \textbf{2008}, \emph{4}, 435--447\relax
\mciteBstWouldAddEndPuncttrue
\mciteSetBstMidEndSepPunct{\mcitedefaultmidpunct}
{\mcitedefaultendpunct}{\mcitedefaultseppunct}\relax
\EndOfBibitem
\bibitem[Ufimtsev and Mart\'{i}nez(2009)Ufimtsev, and
  Mart\'{i}nez]{Ufimtsev2009}
Ufimtsev,~I.; Mart\'{i}nez,~T.~J. Quantum Chemistry on Graphical Processing
  Units. 3. Analytical Energy Gradients and First Principles Molecular
  Dynamics. \emph{J. Chem. Theory Comput.} \textbf{2009}, \emph{5},
  2619--2628\relax
\mciteBstWouldAddEndPuncttrue
\mciteSetBstMidEndSepPunct{\mcitedefaultmidpunct}
{\mcitedefaultendpunct}{\mcitedefaultseppunct}\relax
\EndOfBibitem
\bibitem[Titov \latin{et~al.}(2013)Titov, Ufimtsev, Luehr, and
  Mart\'{i}nez]{Titov2013}
Titov,~A.; Ufimtsev,~I.; Luehr,~N.; Mart\'{i}nez,~T.~J. Generating Efficient
  Quantum Chemistry Codes for Novel Architectures. \emph{J. Chem. Theory
  Comput.} \textbf{2013}, \emph{9}, 213--221\relax
\mciteBstWouldAddEndPuncttrue
\mciteSetBstMidEndSepPunct{\mcitedefaultmidpunct}
{\mcitedefaultendpunct}{\mcitedefaultseppunct}\relax
\EndOfBibitem
\bibitem[Pandya \latin{et~al.}(2022)Pandya, Ashoka, Georgiou, Sung,
  Jayaprakash, Renken, Gai, Shen, Rao, and Musser]{Pandya2022}
Pandya,~R.; Ashoka,~A.; Georgiou,~K.; Sung,~J.; Jayaprakash,~R.; Renken,~S.;
  Gai,~L.; Shen,~Z.; Rao,~A.; Musser,~A.~J. Tuning the Coherent Propagation of
  Organic Exciton-Polaritons through Dark State Delocalization. \emph{Adv.
  Sci.} \textbf{2022}, 2105569\relax
\mciteBstWouldAddEndPuncttrue
\mciteSetBstMidEndSepPunct{\mcitedefaultmidpunct}
{\mcitedefaultendpunct}{\mcitedefaultseppunct}\relax
\EndOfBibitem
\bibitem[Balasubrahmaniyam \latin{et~al.}(2023)Balasubrahmaniyam, Simkhovich,
  Golombek, Sandik, Ankonina, and Schwartz]{Schwartz2022}
Balasubrahmaniyam,~M.; Simkhovich,~A.; Golombek,~A.; Sandik,~G.; Ankonina,~G.;
  Schwartz,~T. From enhanced diffusion to ultrafast ballistic motion of hybrid
  light–matter excitations. \emph{Nature Materials} \textbf{2023},
  \emph{22}\relax
\mciteBstWouldAddEndPuncttrue
\mciteSetBstMidEndSepPunct{\mcitedefaultmidpunct}
{\mcitedefaultendpunct}{\mcitedefaultseppunct}\relax
\EndOfBibitem
\end{mcitethebibliography}


\begin{thebibliography}{10}
\providecommand{\url}[1]{\texttt{#1}}
\providecommand{\urlprefix}{URL }

\bibitem{Freixanet2000}
T.~Freixanet, B.~Sermage, A.~Tiberj, R.~Planel,
\newblock \emph{Phys. Rev. B} \textbf{2000}, \emph{61} 7233.

\bibitem{Coles2014}
D.~M. Coles, N.~Somaschi, P.~Michetti, C.~Clark, P.~G. Lagoudakis, P.~G.
  Savvidis, D.~G. Lidzey,
\newblock \emph{Nature Materials} \textbf{2014}, \emph{13} 712.

\bibitem{Zhong2016}
X.~Zhong, T.~Chervy, S.~Wang, J.~George, A.~Thomas, J.~A. Hutchison, E.~Devaux,
  C.~Genet, T.~W. Ebbesen,
\newblock \emph{Angew. Chem. Int. Ed.} \textbf{2016}, \emph{55} 6202.

\bibitem{Zhong2017}
X.~Zhong, T.~Chervy, L.~Zhang, A.~Thomas, J.~George, C.~Genet, J.~A. Hutchison,
  T.~W. Ebbesen,
\newblock \emph{Angew. Chem. Int. Ed.} \textbf{2017}, \emph{56} 9034.

\bibitem{Lerario2017}
G.~Lerario, D.~Ballarini, A.~Fieramosca, A.~Cannavale, A.~Genco, F.~Mangione,
  S.~Gambino, L.~Dominici, M.~D. Giorgi, G.~Gigli, D.~Sanvitto,
\newblock \emph{Light Sci. Appl.} \textbf{2017}, \emph{6} e16212.

\bibitem{Myers2018}
D.~M. Myers, S.~Mukherjee, J.~Beaumariage, D.~W. Snoke,
\newblock \emph{Phys. Rev. B} \textbf{2018}, \emph{98} 235302.

\bibitem{Rozenman2018}
G.~G. Rozenman, K.~Akulov, A.~Golombek, T.~Schwartz,
\newblock \emph{ACS Photonics} \textbf{2018}, \emph{5} 105.

\bibitem{Zakharko2018}
Y.~Zakharko, M.~Rother, A.~Graf, B.~H\"{a}hnlein, M.~Brohmann, J.~Pezoldt,
  J.~Zaumseil,
\newblock \emph{Nano Lett.} \textbf{2018}, \emph{18} 4927.

\bibitem{Georgiou2018}
K.~Georgiou, P.~Michetti, L.~Gai, M.~Cavazzini, Z.~Shen, D.~G. Lidzey,
\newblock \emph{ACS Photonics} \textbf{2018}, \emph{5} 258.

\bibitem{Xiang2020}
B.~Xiang, R.~F. Ribeiro, M.~Du, L.~Chen, Z.~Yang, J.~Wang, J.~Yuen-Zhou,
  W.~Xiong,
\newblock \emph{Science} \textbf{2020}, \emph{368} 665.

\bibitem{Forrest2020}
S.~Hou, M.~Khatoniar, K.~Ding, Y.~Qu, A.~Napolov, V.~M. Menon, S.~R. Forrest,
\newblock \emph{Adv. Mater.} \textbf{2020}, \emph{32(28)} 2002127.

\bibitem{Pandya2021}
R.~Pandya, R.~Y.~S. Chen, Q.~Gu, J.~Sung, C.~Schnedermann, O.~S. Ojambati,
  R.~Chikkaraddy, J.~Gorman, G.~Jacucci, O.~D. Onelli, T.~Willhammar, D.~N.
  Johnstone, S.~M. Collins, P.~A. Midgley, F.~Auras, T.~Baikie, R.~Jayaprakash,
  F.~Mathevet, R.~Soucek, M.~Du, A.~M. Alvertis, A.~Ashoka, S.~Vignolini, D.~G.
  Lidzey, J.~J. Baumberg, R.~H. Friend, T.~Barisien, L.~Legrand, A.~W. Chin,
  J.~Yuen-Zhou, S.~K. Saikin, P.~Kukura, A.~J. Musser, A.~Rao,
\newblock \emph{Nat. Commun.} \textbf{2021}, \emph{12} 6519.

\bibitem{Georgiou2021}
K.~Georgiou, R.~Jayaprakash, A.~Othonos, D.~G. Lidzey,
\newblock \emph{Angew. Chem. Int. Ed.} \textbf{2021}, \emph{60} 16661.

\bibitem{Berghuis2022}
M.~A. Berghuis, R.~H. Tichauer, L.~de~Jong, I.~Sokolovskii, P.~Bai,
  M.~Ramezani, S.~Murai, G.~Groenhof, J.~G\'{o}mez-Rivas,
\newblock \emph{ACS Photonics} \textbf{2022}, \emph{9} 123.

\bibitem{Pandya2022}
R.~Pandya, A.~Ashoka, K.~Georgiou, J.~Sung, R.~Jayaprakash, S.~Renken, L.~Gai,
  Z.~Shen, A.~Rao, A.~J. Musser,
\newblock \emph{Adv. Sci.} \textbf{2022}, 2105569.

\bibitem{Torma2015}
P.~T{\"{o}}rm{\"{a}}, W.~L. Barnes,
\newblock \emph{Rep. Prog. Phys.} \textbf{2015}, \emph{78} 013901.

\bibitem{Forn-Diaz2019}
P.~Forn-D\'{i}az, L.~Lamata, E.~Rico, J.~Kono, E.~Solano,
\newblock \emph{Rev. Mod. Phys.} \textbf{2019}, \emph{91} 025005.

\bibitem{Rider2022}
M.~S. Rider, W.~L. Barnes,
\newblock \emph{Contemporary Physics} \textbf{2022}, \emph{62}, 4 217.

\bibitem{Agranovich2007}
V.~Agranovich, Y.~Gartstein,
\newblock \emph{Phys. Rev. B} \textbf{2007}, \emph{75} 075302.

\bibitem{Litinskaya2008}
M.~Litinskaya,
\newblock \emph{Phys. Lett. A} \textbf{2008}, \emph{372} 3898.

\bibitem{Schwartz2022}
M.~Balasubrahmaniyam, A.~Simkhovich, A.~Golombek, G.~Sandik, G.~Ankonina,
  T.~Schwartz,
\newblock \emph{Nature Materials} \textbf{2023}, \emph{22}.

\bibitem{Michetti2008}
P.~Michetti, G.~C. La~Rocca,
\newblock \emph{{Phys. Rev. B}} \textbf{{2008}}, \emph{{77}}, {19} 195301.

\bibitem{Allard2022}
T.~F. Allard, G.~Weick,
\newblock \emph{Phys. Rev. B} \textbf{2022}, \emph{106} 245424.

\bibitem{Engelhardt2022}
G.~Engelhardt, J.~Cao,
\newblock \emph{Phys. Rev. B} \textbf{2022}, \emph{105} 064205.

\bibitem{Sokolovskii2022}
I.~Sokolovskii, R.~H. Tichauer, J.~Feist, G.~Groenhof,
\newblock \emph{arXiv} \textbf{2022}, 2209.07309.

\bibitem{Tropf2017}
L.~Tropf, C.~P. Dietrich, S.~Herbst, A.~L. Kanibolotsky, P.~J. Skabara,
  F.~W\"{u}rthner, I.~D.~W. Samuel, M.~C. Gather, S.~Hoefling,
\newblock \emph{Appl. Phys. Lett.} \textbf{2017}, \emph{110} 153302.

\bibitem{Michetti2005}
P.~Michetti, G.~C.~L. Rocca,
\newblock \emph{Phys. Rev. B.} \textbf{2005}, \emph{71} 115320.

\bibitem{Tichauer2021}
R.~Tichauer, J.~Feist, G.~Groenhof,
\newblock \emph{J. Chem. Phys} \textbf{2021}, \emph{154} 104112.

\bibitem{Warshel1976b}
A.~Warshel, M.~Levitt,
\newblock \emph{J. Mol. Biol.} \textbf{1976}, \emph{103} 227.

\bibitem{Boggio-Pasqua2012}
M.~Boggio-Pasqua, C.~F. Burmeister, M.~A. Robb, G.~Groenhof,
\newblock \emph{Phys. Chem. Chem. Phys.} \textbf{2012}, \emph{14} 7912.

\bibitem{Groenhof2019}
G.~Groenhof, C.~Climent, J.~Feist, D.~Morozov, J.~J. Toppari,
\newblock \emph{J. Chem. Phys. Lett.} \textbf{2019}, \emph{10} 5476.

\bibitem{Tichauer2022}
R.~H. Tichauer, D.~Morozov, I.~Sokolovskii, J.~J. Toppari, G.~Groenhof,
\newblock \emph{J. Phys. Chem. Lett.} \textbf{2022}, \emph{13} 6259.

\bibitem{Renken2021}
S.~Renken, R.~Pandya, K.~Georgiou, R.~Jayaprakash, L.~Gai, Z.~Shen, D.~G.
  Lidzey, A.~Rao, A.~J. Musser,
\newblock \emph{J. Chem. Phys.} \textbf{2021}, \emph{155}, 15 154701.

\bibitem{Houdre1996}
R.~Houdr{\'{e}}, R.~P. Stanley, M.~Ilegems,
\newblock \emph{Phys. Rev. A} \textbf{1996}, \emph{53} 2711.

\bibitem{Eizner2019}
E.~Eizner, L.~A. Mart\'{i}nez-Mart\'{i}nez, J.~{Yuen-Shou}, S.~K\'{e}na-Cohen,
\newblock \emph{Sci. Adv.} \textbf{2019}, \emph{5} eaax4484.

\bibitem{Martinez2019}
L.~A. Mart\'{i}nez-Mart\'{i}nez, E.~Eizner, S.~K\'{e}na-Cohen, J.~Yuen-Zhou,
\newblock \emph{J. Chem. Phys.} \textbf{2019}, \emph{151} 054106.

\bibitem{Ehrenfest1927}
P.~Ehrenfest,
\newblock \emph{Z. Phys.} \textbf{1927}, \emph{45} 445.

\bibitem{Luk2017}
H.-L. Luk, J.~Feist, J.~J. Toppari, G.~Groenhof,
\newblock \emph{J. Chem. Theory Comput.} \textbf{2017}, \emph{13} 4324.

\bibitem{Ditchfield1971}
R.~Ditchfield, W.~J. Hehre, J.~A. Pople,
\newblock \emph{J. Chem. Phys.} \textbf{1971}, \emph{54} 724.

\bibitem{Duan2003}
Y.~Duan, C.~Wu, S.~Chowdhury, M.~C. Lee, G.~M. Xiong, W.~Zhang, R.~Yang,
  P.~Cieplak, R.~Luo, T.~Lee, J.~Caldwell, J.~M. Wang, P.~Kollman,
\newblock \emph{J. Comput. Chem.} \textbf{{2003}}, \emph{{24}}, {16} \textbf{1999}.

\bibitem{Jorgensen1983}
W.~L. Jorgensen, J.~Chandrasekhar, J.~D. Madura, R.~W. Impey, M.~L. Klein,
\newblock \emph{J. Chem. Phys.} \textbf{1983}, \emph{79} 926.

\bibitem{Granucci2001}
G.~Granucci, M.~Persico, A.~Toniolo,
\newblock \emph{J. Chem. Phys.} \textbf{2001}, \emph{114} 10608.

\bibitem{Hess2008}
B.~Hess, C.~Kutzner, D.~van~der Spoel, E.~Lindahl,
\newblock \emph{J. Chem. Theory Comput.} \textbf{2008}, \emph{4} 435.

\bibitem{g16}
M.~J. Frisch, G.~W. Trucks, H.~B. Schlegel, G.~E. Scuseria, M.~A. Robb, J.~R.
  Cheeseman, G.~Scalmani, V.~Barone, G.~A. Petersson, H.~Nakatsuji, X.~Li,
  M.~Caricato, A.~V. Marenich, J.~Bloino, B.~G. Janesko, R.~Gomperts,
  B.~Mennucci, H.~P. Hratchian, J.~V. Ortiz, A.~F. Izmaylov, J.~L. Sonnenberg,
  D.~Williams-Young, F.~Ding, F.~Lipparini, F.~Egidi, J.~Goings, B.~Peng,
  A.~Petrone, T.~Henderson, D.~Ranasinghe, V.~G. Zakrzewski, J.~Gao, N.~Rega,
  G.~Zheng, W.~Liang, M.~Hada, M.~Ehara, K.~Toyota, R.~Fukuda, J.~Hasegawa,
  M.~Ishida, T.~Nakajima, Y.~Honda, O.~Kitao, H.~Nakai, T.~Vreven,
  K.~Throssell, J.~A. Montgomery, Jr., J.~E. Peralta, F.~Ogliaro, M.~J.
  Bearpark, J.~J. Heyd, E.~N. Brothers, K.~N. Kudin, V.~N. Staroverov, T.~A.
  Keith, R.~Kobayashi, J.~Normand, K.~Raghavachari, A.~P. Rendell, J.~C.
  Burant, S.~S. Iyengar, J.~Tomasi, M.~Cossi, J.~M. Millam, M.~Klene, C.~Adamo,
  R.~Cammi, J.~W. Ochterski, R.~L. Martin, K.~Morokuma, O.~Farkas, J.~B.
  Foresman, D.~J. Fox,
\newblock {Gaussian 16 revision c.01} \textbf{2016}.

\end{thebibliography}

\end{document}

\setlength{\marginparwidth}{3cm} 
\newpage
\tableofcontents

\section{Molecular dynamics in the collective strong coupling regime}

%light-matter interactions

\subsection{Multi-scale Tavis-Cummings model}

To model the dynamics of $N$ dye molecules strongly coupled to $n_\text{mode}$ confined light modes of a one-dimensional (1D) Fabry-P\'{e}rot cavity, we extended the Tavis-Cummings model\cite{Jaynes1963,Tavis1969} to account for both the molecular degrees of freedom,\cite{Luk2017} and the cavity mode structure:\cite{Tichauer2021}
\begin{equation}
\begin{array}{ccl}
    {\hat{H}}^{\text{TC}} &=& \sum_j^N h\nu_j(\mathbf R_j)\hat{\sigma}^+_j\hat{\sigma}^-_j+\sum_{k_z}^{n_\text{mode}}\hbar\omega_{\text{cav}}(k_z)\hat{a}_{k_z}^\dagger\hat{a}_{k_z}+\\
    \\
    &&\sum_j^N\sum_{k_z}^{n_\text{mode}}\hbar g_j(k_z) \left(\hat{\sigma}^+_j\hat{a}_{k_z}e^{ik_zz_j}+\hat{\sigma}^-_j\hat{a}_{k_z}^\dagger e^{-ik_zz_j}\right)+\\
    \\
    &&\sum_{i}^N V^\text{mol}_{\text{S}_0}({\bf{R}}_i)
    %+\sum_i^N T({\bf{P}}_j)
    \end{array}
    \label{eq:dTCH}
\end{equation}
Here, $\hat{\sigma}_j^+$ ($\hat{\sigma}_j^-$) is the operator that excites (de-excites) molecule $j$ from the electronic ground (excited) state $|\text{S}_0^j({\bf{R}}_j)\rangle$ ($|\text{S}_1^{j}({\bf{R}}_j)\rangle$) into the electronic excited (ground) state $|\text{S}_1^{j}({\bf{R}}_j)\rangle$ ($|\text{S}_0^j({\bf{R}}_j)\rangle$); ${\bf{R}}_j$ is the vector of the Cartesian coordinates of all atoms in molecule $j$, centered at $z_j$;
$\hat{a}_{k_z}$ ($\hat{a}_{k_z}^\dagger$) is the annihilation (creation) operator of an excitation of a cavity mode with wave-vector $k_z$; $h\nu_j(\mathbf R_j)$ is the excitation energy of molecule $j$, defined as: 
\begin{equation}
  h\nu_j(\mathbf R_j)=V_{\text{S}_1}^{\text{mol}}({\bf{R}}_j)-V_{\text{S}_0}^{\text{mol}}({\bf{R}}_j)
  \label{eq:Hnu}
\end{equation}
with $V_{\text{S}_0}^{\text{mol}}({\bf{R}}_j)$ and $V_{\text{S}_1}^{\text{mol}}({\bf{R}}_j)$ the adiabatic potential energy surfaces of molecule $j$ in the electronic ground (S$_0$) and excited (S$_1$) state, respectively. 

The last term in Equation~\ref{eq:dTCH} is the total potential energy of the system in the absolute ground state ({\it i.e.}, with no excitations in neither the molecules nor the cavity modes), defined as the sum of the ground-state potential energies of all molecules in the cavity. The $V_{\text{S}_0}^{\text{mol}}({\bf{R}}_j)$ and $V_{\text{S}_1}^{\text{mol}}({\bf{R}}_j)$ adiabatic potential energy surfaces are modelled at the QM/MM level of theory,\cite{Warshel1976b,Boggio-Pasqua2012} as described in the Computational Details section of the main text.

\begin{figure*}[!htb]
\centerline{
\epsfig{figure=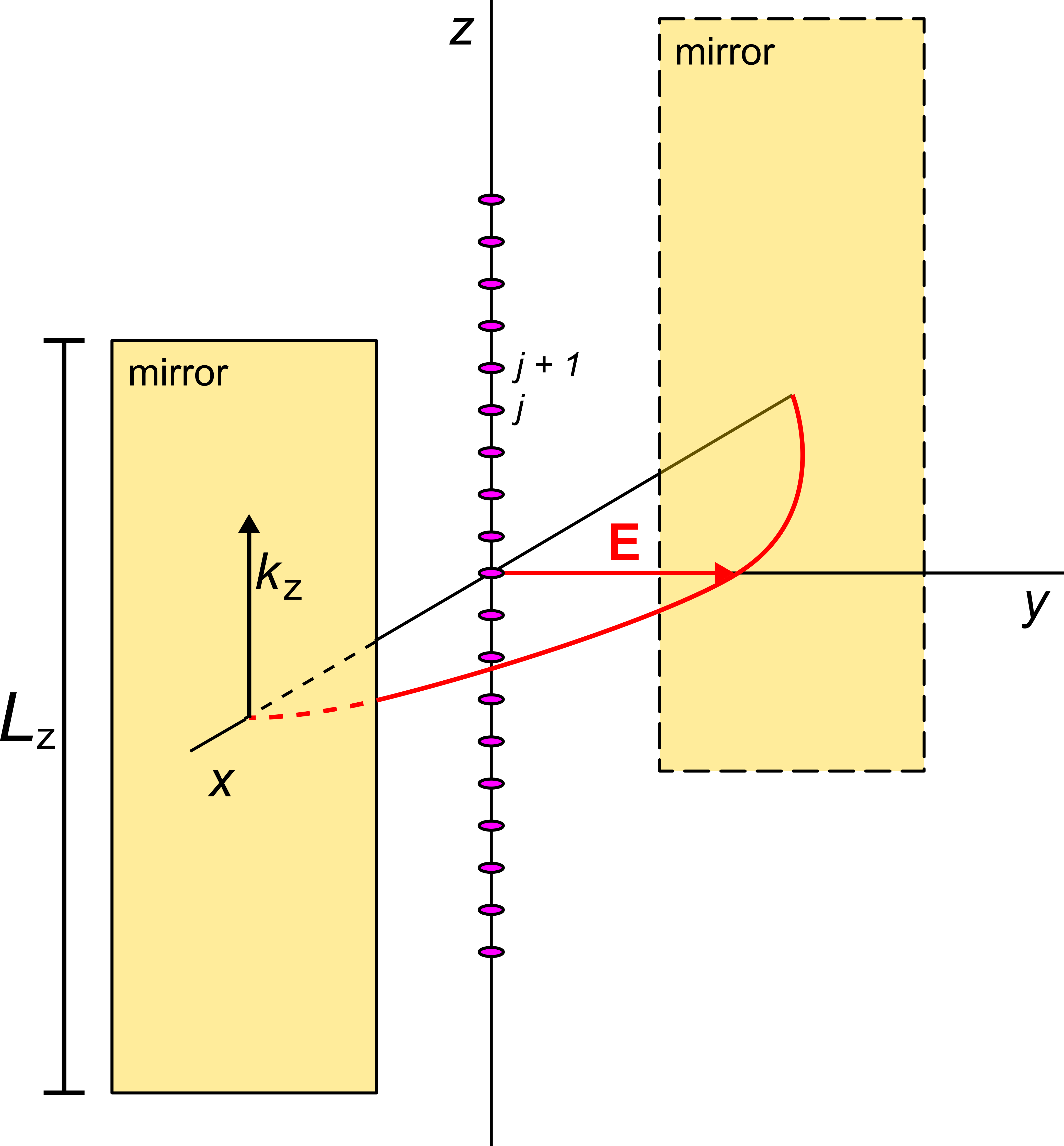,width=7cm,angle=0}}
  \caption{One-dimensional (1D) Fabry-P\'{e}rot micro-cavity model.\cite{Michetti2005} Two reflecting mirrors located at $-\frac{1}{2}x$ and $\frac{1}{2}x$, confine light modes along this direction, while free propagation along the $z$ direction is possible for plane waves with in-plane momentum $k_z$ and energy $\hslash\omega_{\text{cav}}(k_z)$. The vacuum field vector (red) points along the $y$-axis, reaching a maximum amplitude at $x=0$ where the $N$ molecules (magenta ellipses) are placed, distributed along the $z$-axis at positions $z_j$ with $1\le j \le N$. 
  }
  \label{fig:1D_cavity}
\end{figure*}

The third term in Equation~\ref{eq:dTCH} models the light-matter interaction within the dipolar approximation through $g_j(k_z)$:
\begin{equation}
    g_j(k_z) = -{\boldsymbol{\mu}}^\text{TDM}_j({\bf{R}}_j) \cdot {\bf{u}}_{\text{cav}} \sqrt{\frac{\hbar\omega_{\text{cav}}(k_z)}{2\epsilon_0 V_{\text{cav}}}} 
    \label{eq:dipole_coupling}
\end{equation}
where ${\boldsymbol{\mu}}_j^\text{TDM}({\bf{R}}_j)$ is the transition dipole moment of molecule $j$ that depends on the molecular geometry (${\bf{R}}_j)$; ${\bf{u}}_\text{cav}$ the unit vector in the direction of the electric component of the cavity vacuum field ({\it{i.e.}}, $|{\bf{E}}|=\sqrt{\hbar\omega_\text{cav}(k_z)/2\epsilon_0V_\text{cav}}$), chosen along the $y$-direction (see Figure~\ref{fig:1D_cavity}); $\epsilon_0$ the vacuum permittivity; and $V_{\text{cav}}$ the cavity mode volume.

% figure 1S

\subsection{Multi-mode cavity model}\label{subsec:cavity}

To discretize the cavity dispersion, we follow Michetti and La Rocca,\cite{Michetti2005} and impose periodic boundary conditions in the $z$-direction of the 1D cavity. Under these conditions, the wave vectors, $k_z$, adopt discrete values: $k_{z,p}=2\pi p/L_z$ with $p \in \mathbb Z$ and $L_{z}$ the length of the 1D cavity. After discretization the molecular Tavis-Cummings Hamiltonian in Equation~\ref{eq:dTCH} becomes a $(N+n_{\text{mode}})$ by $(N+n_{\text{mode}})$ matrix with four blocks:\cite{Tichauer2021}
\begin{equation}
  {\bf{H}}^{\text{TC}} = \left(\begin{array}{cc} 
  {\bf{H}}^{\text{mol}} & {\bf{H}}^{\text{int}}\\
  {\bf{H}}^{\text{int}\dagger} & {\bf{H}}^{\text{cav}}\\
  \end{array}\right)\label{eq:TavisCummings}
\end{equation}
The elements of this matrix are evaluated in the product basis of adiabatic molecular states times cavity mode excitations:
\begin{equation}
\begin{array}{ccl}
|\phi_j\rangle &=& \hat{\sigma}_j^+|\text{S}_0^1\text{S}_0^2..\text{S}_0^{N-1}\text{S}_0^N\rangle\otimes|00..0\rangle\\
\\
&=&\hat{\sigma}_j^+|\Pi_i^N\text{S}_0^i\rangle\otimes|\Pi_k^{n_\text{mode}}0_k\rangle \\
\\
&=&\hat{\sigma}_j^+|\phi_0\rangle
\end{array}
\end{equation}
for $1\le j\le N$, and 
\begin{equation}
\begin{array}{ccl}
|\phi_{j >N}\rangle &=& \hat{a}_{j-N}^\dagger|\text{S}_0^1\text{S}_0^2..\text{S}_0^{N-1}\text{S}_0^N\rangle\otimes|00..0\rangle\\
\\
&=&\hat{a}_{j-N}^\dagger|\Pi_i^N\text{S}_0^i\rangle\otimes|\Pi_k^{n_\text{mode}}0_k\rangle \\
\\
&=&\hat{a}_{j-N}^\dagger|\phi_0\rangle
\end{array}
\end{equation}
for $N < j\le N+n_\text{mode}$. In these expressions $|00..0\rangle$ indicates that all Fock states associated with the $n_\text{mode}$ cavity modes are empty. The basis state $|\phi_0\rangle$ is the ground state of the molecule-cavity system with no excitations in neither the molecules nor cavity modes:
\begin{equation}
|\phi_0\rangle = |\text{S}_0^1\text{S}_0^2..\text{S}_0^{N-1}\text{S}_0^N\rangle\otimes|00..0\rangle
=|\Pi_i^N\text{S}_0^i\rangle\otimes|\Pi_k^{n_\text{mode}}0_k\rangle\label{eq:phi0}
\end{equation}

The upper left block, ${\bf{H}}^{\text{mol}}$, is an $N\times N$ matrix that contains the single-photon excitations of the molecules. Because we neglect direct excitonic interactions between molecules, this block is diagonal, with elements labeled by the molecule indices $j$:
\begin{equation}
    H_{j,j}^{\text{mol}}=\langle \phi_0|\hat{\sigma}_j\hat{H}^{\text{TC}}\hat{\sigma}_j^+|\phi_0\rangle\label{eq:molecular_diagonal}
\end{equation}
for $1 \le j \le N$. Each matrix element of ${\bf{H}}^\text{mol}$ thus represents the potential energy of a molecule, $j$, in the electronic excited state $|\text{S}_1^{j}({\bf{R}}_j)\rangle$ while all other molecules, $i\neq j$, are in the electronic ground state $|\text{S}_0^i({\bf{R}}_i)\rangle$:
\begin{equation}  H_{j,j}^{\text{mol}}=V_{\text{S}_1}^{\text{mol}}({\bf{R}}_j)+\sum^N_{i\neq j}V_{\text{S}_0}^{\text{mol}}({\bf{R}}_i)
  \label{eq:Hj}
\end{equation}

The lower right block in Equation~\ref{eq:TavisCummings},  ${\bf{H}}^{\text{cav}}$, is an $n_\text{mode}\times n_\text{mode}$ matrix (with $n_{\text{mode}}=n_{\text{max}}-n_{\text{min}}+1$) containing the single-photon excitations of the cavity modes, and is also diagonal:
\begin{equation}
    H_{p,p}^{\text{cav}}= \langle \phi_0|\hat{a}_p\hat{H}^{\text{TC}}\hat{a}^\dagger_p|\phi_0\rangle\label{eq:photonic_diagonal}
\end{equation}
for $n_{\text{min}}\le p \le n_{\text{max}}$. 
Here, $\hat{a}_p^\dagger$ excites cavity mode $p$ with wave-vector $k_{z,p}= 2\pi p/L_z$. In these matrix elements, all molecules are in the electronic ground state (S$_0$). The energy is therefore the sum of the cavity energy at $k_{z,p}$, and the molecular ground state energies:
\begin{equation}
    H_{p,p}^{\text{cav}}= \hslash\omega_{\text{cav}}(2\pi p/L_z)+\sum^N_{j}V_{\text{S}_0}^{\text{mol}}({\bf{R}}_j)
    \label{eq:Hn}
\end{equation}
where, $\omega_{\text{cav}}(k_{z,p})$ is the cavity dispersion (dashed-dotted curve in Figure 1{\bf{b}}, main text): 
\begin{equation}
  \omega_\text{cav}(k_{z,p})=\sqrt{\omega^2_0+ c^2k_{z,p}^2/n^2}
  \label{eq:dispersion}
\end{equation}
with $\hslash\omega_0$ the energy at $k_{z,0}=0$, $n$ the refractive index of the medium and $c$ the speed of light in vacuum.

The two $N\times n_\text{mode}$ off-diagonal blocks ${\bf{H}}^{\text{int}}$ and ${\bf{H}}^{\text{int}\dagger}$ in the multi-mode Tavis-Cummings Hamiltonian (Equation~\ref{eq:TavisCummings}) model the light-matter interactions between the molecules and the cavity modes. These matrix elements are approximated as the inner product between the molecular transition dipole moments on the one hand, and the transverse electric field of the cavity modes at the center of the molecules, on the other hand:
\begin{equation}
\begin{array}{ccl}
  H_{j,p}^{\text{int}} &=& 
  -{\boldsymbol{\upmu}}_j^\text{TDM}({\bf{R}}_j)\cdot {\bf{u}}_{\text{cav}} \sqrt{\frac{\hslash\omega_{\text{cav}}(2\pi p/L_z)}{2\epsilon_0 V_{\text{cav}}}} \langle\phi_0|\hat{\sigma}_j(\hat{\sigma}^+_j\hat{a}_p e^{i 2\pi p z_j/L_z})\hat{a}_p^\dagger|\phi_0\rangle \\
  \\
  &=&
  -{\boldsymbol{\upmu}}_j^\text{TDM}({\bf{R}}_j) \cdot {\bf{u}}_{\text{cav}} \sqrt{\frac{\hslash\omega_{\text{cav}}(2\pi p/L_z)}{2\epsilon_0 V_{\text{cav}}}} e^{i 2\pi p z_j/L_z}     
  \end{array}
  \label{eq:QMMMdipole_coupling}
\end{equation}
for $1 \le j \le N$ and $n_{\text{min}}\le p \le n_{\text{max}}$. 

Diagonalization of the multi-scale Tavis-Cummings Hamiltonian in Equation~\ref{eq:TavisCummings} yields the $N+n_\text{mode}$ hybrid light-matter states $\lvert\psi^m\rangle$:\cite{Michetti2005,Agranovich2007}
\begin{equation}
\vert\psi^{m}\rangle=\left(\sum_{j}^N\beta^m_j\hat{\sigma}^+_j + \sum_{p}^{n_{\text{mode}}}\alpha^m_p\hat{a}_p^\dagger\right)\vert\phi_0\rangle\label{eq:Npolariton}
\end{equation}
with eigenenergies $E_m$. The expansion coefficients $\beta_j^m$ and $\alpha_p^m$ reflect , respectively, the contribution of the molecular excitons ($\vert\text{S}_1^j(\mathbf{R}_j)\rangle$) and the cavity mode excitations ($\vert 1_p\rangle$) to polariton $\vert \psi^m\rangle$. 

\subsection{Ehrenfest molecular dynamics simulations}

MD trajectories of all molecules (including environment) were computed by numerically integrating Newton's equations of motion. The multi-mode Tavis-Cummings Hamiltonian (Equation~\ref{eq:TavisCummings}) was diagonalized at each time-step of the simulation to obtain the $N+n_\text{mode}$ (adiabatic) polaritonic eigenstates $|\psi^m\rangle$ and energies $E^m$.
The \emph{total} polaritonic wavefunction $|\Psi(t)\rangle$ was coherently propagated along with the classical degrees of freedom of the $N$ molecules as a time-dependent superposition of the $N+n_\text{mode}$ time-independent adiabatic polaritonic states:
\begin{equation}
|\Psi(t)\rangle=\sum_m^{N+n_{\text{mode}}}c_m(t)|\psi ^m\rangle\label{eq:totalwf}
\end{equation}
where $c_m(t)$ are the time-dependent expansion coefficients of the time-independent eigenstates, $|\psi^m\rangle$, defined in Equation~\ref{eq:Npolariton}. A unitary propagator in the \emph{local} diabatic basis was used to integrate these coefficients,\cite{Granucci2001} while the nuclear degrees of freedom of the $N$ molecules evolved on the mean-field potential energy surface:
\begin{equation}
V({\bf{R}})=\langle\Psi(t)|\hat{H}^\text{TC}|\Psi(t)\rangle
\end{equation}

 \subsection{Cavity decay}

%cavity decay
Radiative loss through the imperfect cavity mirrors was modeled as a first-order decay process into the overall ground state of the system ({\it{i.e.}}, no excitation in neither the molecules nor the cavity modes).~\cite{Luk2017} Assuming that the intrinsic decay rates, $\gamma_{\text{cav}}$, are the same for all modes, the total loss rate was calculated as the product of $\gamma_{\text{cav}}$ and the total photonic weight, $\sum_{p}^{n_\text{mode}}|\alpha^m_p|^2$, of state $|\psi^m\rangle$. Thus, after an MD step $\Delta t$, the population in state $|\psi_m\rangle$, $\rho_m(t)= |c_m(t)|^2$, becomes:
\begin{equation}
  \rho_m(t+\Delta t) = \rho_m(t) \exp\left[-\gamma_{\text{cav}}\sum_p^{n_\text{mode}}|\alpha^m_p(t)|^2\Delta t\right]\label{eq:decay_tot}
\end{equation}
Since $\rho_m = (\Re[c_m])^2+(\Im[c_m])^2$, changes in the real and imaginary parts of the  (complex) expansion coefficients $c_m(t)$ due to spontaneous photonic loss through the mirrors of a low-Q cavity, are:
\begin{equation*}
    \Re[c_m (t+\Delta t)] = \Re[c_m
                              (t)] \exp\left[-\frac{1}{2}\gamma_{\text{cav}}\sum_p^{n_\text{mode}}|\alpha^m_p(t)|^2\Delta
                              t\right]
\end{equation*}
\begin{equation*}
 \Im[c_m (t+\Delta t)] = \Im[c_m
                              (t)] \exp\left[-\frac{1}{2}\gamma_{\text{cav}}\sum_p^{n_\text{mode}}|\alpha_p^m(t)|^2\Delta
                              t\right]
\end{equation*}
Simultaneously, the population of the zero-excitation subspace, or ground state, $\rho_0(t+\Delta t)$, increases as:
\begin{equation}
\rho_0(t+\Delta t)=\rho_0(t)+\sum_m\rho_m(t)\left(1-\exp\left[-\gamma_{\text{cav}}\sum_p^{n_\text{mode}}|\alpha_p^m(t)|^2\Delta t\right]\right)
\end{equation}

\section{Further simulation details}

\subsection{Rhodamine model}

\begin{figure*}[!htb]
\centerline{ \epsfig{figure=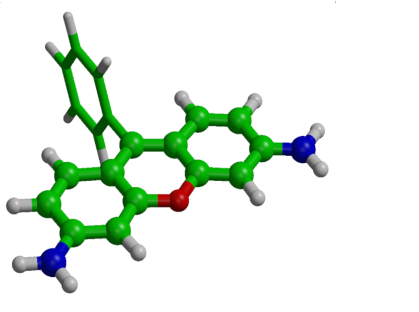,width=8cm,angle=0}}
  \caption{Rhodamine model used in our simulations. The QM subsystem, shown in ball-and-stick representation, is described at the HF/3-21G level of theory in the electronic ground state (S$_0$), and at the CIS/3-21G level of theory in the first singlet excited state (S$_1$). The MM subsystem, consisting of the atoms shown in stick representation, and the water molecules (not shown), are modelled with the Amber03 force field.}
  \label{fig:rhodamine}
\end{figure*}

The Rhodamine molecules, one of which is shown schematically in Figure~\ref{fig:rhodamine}, were modelled with the Amber03 force field,\cite{Duan2003} using the parameters provided by Luk {\it et al.}\cite{Luk2017} After a geometry optimization at the force field level, the molecule was placed at the center of a rectangular box and 3,684 TIP3P water molecules,\cite{Jorgensen1983} were added. The simulation box, which contained 11,089 atoms, was equilibrated for 2~ns with harmonic restraints on the heavy atoms of Rhodamine (force constant 1000~kJmol$^{-1}$nm$^{-1}$). Subsequently, a 200~ns classical MD trajectory was computed at constant temperature (300~K) using a stochastic dynamics integrator with a friction coefficient of 0.1~ps$^{-1}$. The pressure was kept constant at 1 bar using the Berendsen isotropic pressure coupling algorithm\cite{Berendsen1984} with a time constant of 1~ps. The LINCS algorithm was used to constrain bond lengths,\cite{Hess1997} while SETTLE was applied to constrain the internal degrees of freedom of water molecules,\cite{Miyamoto1992} enabling a time step of 2~fs in the classical MD simulations. A 1.0~nm cut-off was used for Van der Waals' interactions, which were modeled with Lennard-Jones potentials. Coulomb interactions were computed with the smooth particle mesh Ewald method,\cite{Essmann1995} using a 1.0~nm real space cut-off and a grid spacing of 0.12~nm. The relative tolerance at the real space cut-off was set to 10$^{-5}$.

% figure 2S

From the second half of the 200~ns MD trajectory, snapshots were extracted and subjected to further equilibration for 10~ps at the QM/MM level. The time step was reduced to 1~fs. As in previous work,\cite{Luk2017,Groenhof2019,Tichauer2021} the fused ring system was included in the QM region and described at the RHF/3-21G level, while the rest of the molecule as well as the water solvent, were modelled with the Amber03 force field\cite{Duan2003} and TIP3P water model,\cite{Jorgensen1983} respectively (Figure~\ref{fig:rhodamine}). The bond connecting the QM and MM subsystems was replaced by a constraint and the QM part was capped with a hydrogen atom. The force on the cap atom was distributed over the two atoms of the bond.\cite{Groenhof2004} The QM system experienced the Coulomb field of all MM atoms within a 1.6~nm cut-off sphere and Lennard-Jones interactions between MM and QM atoms were added. The singlet electronic excited state (S$_1$) was modeled with the Configuration Interaction method, truncated at single electron excitations, for the QM region ({\it {i.e.,}} CIS/3-21G//Amber03). At this level of QM/MM theory, the excitation energy is 4.18~eV~\cite{Luk2017}. The QM/MM simulations were performed with GROMACS version 4.5.3,\cite{Hess2008} interfaced to TeraChem version 1.93.\cite{Ufimtsev2009,Titov2013}

\subsection{Initial conditions}

The initial excitation of a polariton wavepacket was created by assigning to the expansion coefficients $c_m(t=0)$ values of a Gaussian distribution centered at $k$-vector $k_c$~= $80\frac{2\pi}{L_z}$ = 10.05 $\upmu$m$^{-1}$ and covering the whole UP branch ({\it{i.e.}}, excluding LP and dark states in the initial wavepacket):~\cite{Agranovich2007} 

\begin{equation}
    c_m(0) = \left(\frac{\zeta}{2\pi^3}\right)^{\frac{1}{4}}\exp[-\zeta(k_{z}^m-k_c)^2]\label{eq:wavepacket_excitation}
\end{equation}
where $\zeta = 10^{-14}$ $\text{m}^2$ is a coefficient characterising the shape of the wavepacket and $k_{z}^m$ the expectation value of the in-plane momentum of polariton $|\psi_m\rangle$, evaluated as: 
\begin{equation}
    \langle k_{z}^m \rangle = \frac{\sum_p^{n_\text{mode}} |\alpha_p^m|^2 k_{z,p} }{ \sum_p^{n_\text{mode}}|\alpha_p^m|^2}
\end{equation}
with $k_{z,p} = 2\pi p/L_z$ the discrete wave vector in a periodic 1D cavity of length $L_z$ (see subsection~\ref{subsec:cavity}).

\section{Wavepacket analysis}

\subsection{Populations of the lower polariton, upper polariton, and dark states}
\label{sec:mol-pho_parts}

The time evolution of the  populations in the LP, UP, and dark states (plotted in panels {\bf g}, {\bf h}, {\bf i} of Figure~2 in the main text) were obtained by summing over  expansion coefficients that belong to LP, UP or dark states:  {\it i.e.}, $\sum_m|c_m(t)|^2$, with $m \in$~LP (160 low-energy states), UP (160 higher-energy states), or dark states (the remaining states), respectively. With a Rabi splitting of $\sim$ 325~meV, the LP and UP branches are sufficiently separated from the dark state manifold for a correct assignment of all states in our simulations.

\subsection{Wavepacket propagation}
\label{sec:analysis-wps}

To monitor the propagation of the wavepackets, we plotted the probability density of the total time-dependent wave function $|\Psi(t)|^2$ 
at the positions of the molecules,  $z_j$, as a function of time (panels {\bf a}--{\bf f} in figure~2 in the main text). We thus represent the density as a \emph{discrete} distribution at grid points that correspond to the molecular positions, rather than as a continuous distribution. 

The probability density of the total time-dependent wave function $|\Psi(t)|^2$ was calculated as the sum of the probability densities of the molecular $|\Psi_{\text{mol}}(t)|^2$ and photonic $|\Psi_{\text{pho}}(t)|^2$ contributions. The amplitude of $|\Psi_{\text{mol}}(t)\rangle$ at position $z_j$ in the 1D cavity (with $z_j=(j-1)L_z/N$ for $1\le j \le N$) was obtained by projecting the excitonic basis state in which molecule $j$ at position $z_j$ is excited, onto the total wave function (Equation~\ref{eq:totalwf}):
\begin{equation}
\begin{array}{ccl}
    |\Psi^{\text{mol}}(z_j,t)\rangle&=&(\hat{\sigma}_j^+|\phi_0\rangle
        \langle\phi_0|\hat{\sigma}_j)|\Psi(t)\rangle\\
    \\
    &=& \sum_m^{N+n_\text{mode}}c_m(t)\beta_j^m\hat{\sigma}_j^+|\phi_0\rangle
   \end{array}\label{eq:molecularpart}
\end{equation}
with $\beta_j^m$ the expansion coefficient of the excitonic basis state $\sigma_j^+|\phi_0\rangle$ in polaritonic state $|\psi^m\rangle$ (Equation~\ref{eq:Npolariton}), $c_m(t)$ the time-dependent expansion coefficients of the total wavefunction $|\Psi(t)\rangle$ (Equation~\ref{eq:totalwf}), and $|\phi_0\rangle$ the ground state of the molecule-cavity system with no excitations of neither the molecules nor cavity modes (Equation~\ref{eq:phi0}).

The cavity mode excitations are described as plane waves that are delocalized in real space. We therefore obtained the amplitude of the cavity mode excitations in polaritonic eigenstate $|\psi^m\rangle$ at position $z_j$ by Fourier transforming the projection of the cavity mode Fock states, 
%cavity mode excitation basis states, 
in which cavity mode $p$ is excited, onto $|\psi^m\rangle$: %($|\phi_p\rangle=\hat{a}_{p}^\dagger|\phi_0\rangle$):
\begin{equation}
\begin{array}{ccl}
    |\psi^m_\text{pho}(z_j)\rangle&=&\mathcal{FT}^{-1}\left[\sum_p^{n_\text{mode}}(\hat{a}_p^\dagger|\phi_0\rangle\langle\phi_0|\hat{a}_p)|
    \psi^m\rangle\right]\\
    \\
    &=&\frac{1}{\sqrt N}\sum_p^{n_{\text{mode}}} \alpha_p^me^{i2\pi z_j p} \hat{a}_p^\dagger|\phi_0\rangle
    \end{array}\label{eq:photonicpart1}
\end{equation}
where $\alpha_p^m$ is the expansion coefficient of the cavity mode excitation $\alpha_p^{\dagger}|\phi_0\rangle$ in polaritonic state $|\psi^m\rangle$ (Equation~\ref{eq:Npolariton}) and we normalized by $1/\sqrt{N}$ rather than $1/\sqrt{L_z}$, as we represent the density on the grid of molecular positions. The total contribution of the cavity mode excitations to the wavepacket at position $z_j$ at time $t$ was then obtained as the weighted sum over the Fourier transforms:
\begin{equation}
\begin{array}{ccl}
|\Psi^\text{pho}(z_j,t)\rangle &=& \sum_m^{N+n_\text{mode}}c_m(t)\times\mathcal{FT}^{-1}\left[\sum_p^{n_\text{mode}}(\hat{a}_p^\dagger|\phi_0\rangle\langle\phi_0|\hat{a}_p)|\psi^m\rangle\right]\\
\\
&=& \sum_m^{N+n_{\text{mode}}}c_m(t)\frac{1}{\sqrt N}\sum_p^{n_{\text{mode}}} \alpha_p^me^{i2\pi z_jp}\hat{a}_p^\dagger|\phi_0\rangle 
\end{array}\label{eq:photonicpart2}
\end{equation}
with $c_m(t)$ the time-dependent expansion coefficient of the adiabatic polaritonic state $|\psi^m\rangle$ in the total wave function $|\Psi(t)\rangle$ (Equation~\ref{eq:totalwf}).

\subsection{Transient transmission}

\begin{figure*}[!htb]
\centering
\includegraphics[width=12cm]{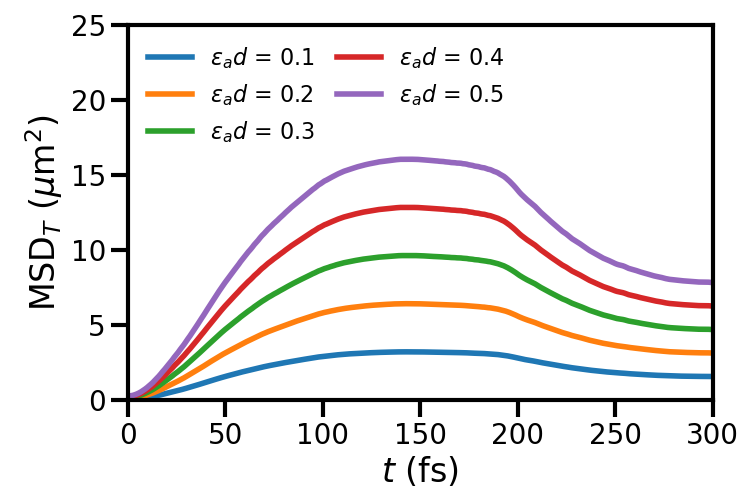}
  \caption{Mean squared displacement (MSD$_T$) of the $\Delta T(z,t)/T_0$ at time $t$ after instantaneous excitation of a Gaussian wavepacket of UP states in a cavity with 512 molecules and a lifetime of $\tau_{\text{cav}}=60$~fs, plotted for various values of $\epsilon_\text{a}d$.
}
\label{fig:MSD_diff_abs_coef}
\end{figure*}

Under the assumption that we can neglect reflection, the transmission of light through the system, $T$, is related to absorbance, $A$, via the Lambert-Beer law:
\begin{equation}
-\ln(T)=A=\varepsilon_\text{a}Cd\label{eq:transmission1}
\end{equation}
with $\varepsilon_{\text{a}}$ the absorption coefficient, $C$ the concentration of absorbers (in our case the Rhodamines), and $d$ the length of the path through which the light passes. Based on the article of Pandya and co-workers,\cite{Pandya2022} we assume that they probed how the transmission of photons with an energy equal to the excitation from the total ground state $|\phi_0\rangle$ (Equation~\ref{eq:phi0}, {\it{i.e.}}, no excitation in neither the molecules, nor cavity) into the LP branch at 640~nm for their BODIPY-R cavity systems, changes as a function of time ($t$) and position (here $z$) after the interaction of the molecule-cavity system with the pump pulse. Assuming furthermore that excitation from the LP or UP into the two-photon manifold is negligible due to absence of resonant transitions at that wavelength, the absorbance of the sample at position $z$ and time $t$ is proportional to the concentration of unexcited molecules:
\begin{equation}
    C(z,t)=C_0-|\Psi(z,t)|^2
\end{equation}
with $C_0$ the concentration of the Rhodamine molecules in the cavity \emph{before} interaction with the pump-pulse.
%The concentration of the ground state at position $z$ and time $t$ $[\rho_0]_{z,t}$ depends on the population of the light-matter hybrid states, including both bright and dark states, {\it{i.e.}},
%\begin{equation}
  %  c(z,t)=c_0-|\Psi(z,t)|^2
%\end{equation}
%where $c_0$ is the concentration of excitable material \emph{before} interaction with the pump-pulse. 
We assume that $C_0$ is uniform and homogeneous. With these approximations, the transient normalized differential transmission in our simulations was calculated as:
\begin{equation}
\begin{array}{ccl}
    \frac{\Delta T(z,t)}{T_0}&=&\frac{T(z,t)-T_0}{T_0}\\
    \\
    &=&\frac{e^{-\varepsilon_{\text{a}}dC({z,t})}-e^{-\varepsilon_{\text{a}}dC_0}}{e^{-\varepsilon_{\text{a}}dC_0}}\\
    \\
    &=&\frac{e^{-\varepsilon_\text{a}d(C_0-|\Psi(z,t)|^2)}-e^{-\varepsilon_\text{a}dC_0 }}{e^{-\varepsilon_\text{a}dC_0 }}\\
    \\
    &=&e^{\varepsilon_\text{a}d|\Psi(z,t)|^2 }-1
    \end{array}\label{eq:transmission2}
\end{equation}
In analogy to Equation S4 in the Supporting Information of the work by Balasubrahmanyam {\it{et al.}}~\cite{Schwartz2022}, the mean squared displacement (MSD) of the transient transmisson signal $\Delta T/T_0$ was calculated as:
\begin{equation}
   \text{MSD}_T = \sum_i^N \left(z_i-z_0\right)^2\frac{\Delta T(z_i,t)}{T_0} = \sum_i^N\left(z_i-z_0\right)^2
   \left[e^{\varepsilon_\text{a}d|\Psi(z_i,t)|^2}-1\right]
\end{equation}
Here, we treated $\varepsilon_\text{a}d$ as a single parameter between 0 and 1. In Figure~\ref{fig:MSD_diff_abs_coef}, we plot the MSD$_T$ of the transmitted signal for various values of $\varepsilon_\text{a}d$. Based on the similarity of the plots, we conclude that the results of the analysis are not very sensitive to the choice of this parameter.

%MSD at different absorption coefficients

\subsection{Estimation of the duration and propagation velocity of the ballistic phase}

\begin{figure*}[!htb]
\centering
\includegraphics[width=16cm]{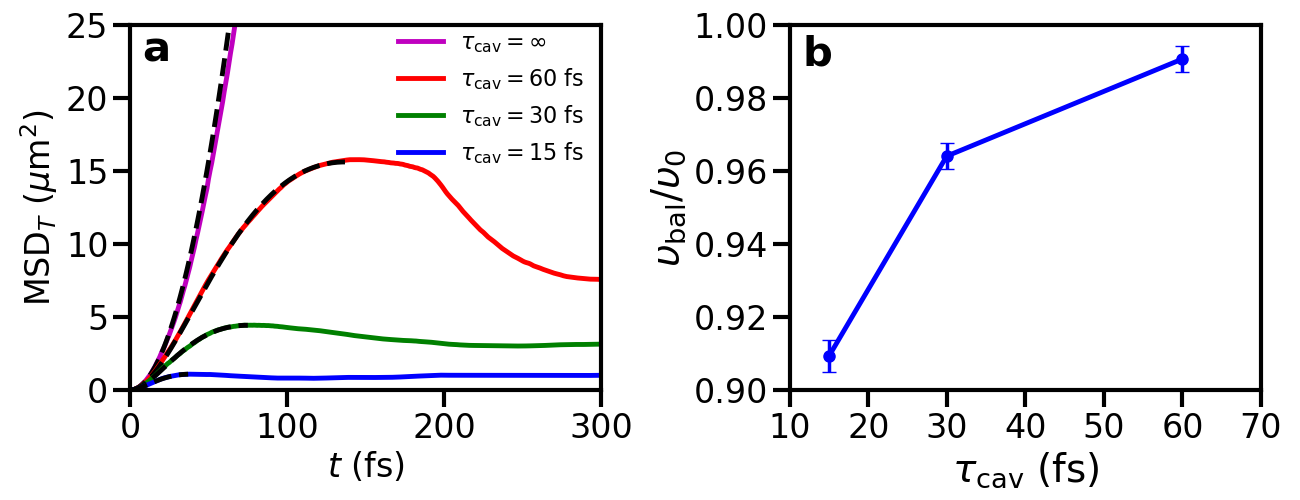}
  \caption{Panel~{\bf {a}}: MSD$_T$ of the transient differential transmission after excitation in the perfect cavity (purple) and in cavities with the decay rates $\tau_{\text{cav}}$ of 60~fs~(red), 30~fs~(green), and 15~fs~(blue). Panel~{\bf {b}}: Average polariton group velocity $\upsilon_{\text{gr}}$ as a function of the cavity lifetime, plotted as a fraction of the group velocity $\upsilon^0_{\text{gr}}$ in the perfect cavity, extracted from a quadratic fit to MSD$_T$s (Equation~\ref{eq:fitting_MSD_trans}).
}
\label{fig:MSD_and_group_vel}
\end{figure*}

The propagation velocity, $\upsilon_{\text{bal}}$, and duration, $\tau_{\text{bal}}$, of the ballistic phase were obtained by fitting the model of Pandya {\it{et al.}} (Equation~5 in their paper~\cite{Pandya2022}) without $\sigma_0$ (which is zero in our simulations) to the MSD${_T}$ of $\Delta T(z,t)/T_0$ from $t=0$ to $t=t_{\text{MSD}_T^{\text{max}}}$, when the MSD$_\text{T}$ reaches its maximum:
\begin{equation}
   \text{MSD}_T(t) = \upsilon_{\text{bal}}^2t^2\exp{(-t/\tau_{\text{bal}})}
   \label{eq:fitting_MSD_trans}
\end{equation}
In Figure~\ref{fig:MSD_and_group_vel}{\bf{a}} these fits
are plotted as dashed lines. The good agreement between the fit and the MSD$_T$ suggests that this function captures the initial MSD$_T$ as a function of time for all cavities.

The propagation velocity, $\upsilon_{\text{bal}}$, only weakly depends on the cavity lifetime, $\tau_{\text{cav}}$ (Figure~\ref{fig:MSD_and_group_vel}{\bf{b}}), because transport in the ballistic regime is primarily governed by the group velocities of UP states. The duration of the ballistic regime, $\tau_{\text{bal}}$, however, depends more strongly on the cavity lifetime (Figure 4{\bf{b}} in the main text), suggesting that the increase of propagation distance with cavity lifetime is mainly due to a longer duration of the ballistic phase.

%Average group velocity from fitting

The error bars in the plots of $\upsilon_{\text{bal}}$ (Figure~\ref{fig:MSD_and_group_vel}{\bf{b}}) and $\tau_{\text{bal}}$ (Figure 4{\bf{b}} in the main text) correspond to the standard deviation $\sigma_x$ of the $S = 5$ trajectories for each cavity mode lifetime:
\begin{equation}
   \sigma_x = \sqrt{\frac{\sum_{i=1}^S \left(x_i-\overline{x}\right)^2}{S-1}}
   \label{eq:stdiv}
\end{equation}
where $x_i$ and $\overline{x}$ are, respectively, the values of 
$\upsilon_{\text{gr}}$ or $\tau_{\text{bal}}$, and their averages, respectively.

\subsection{Estimation of the diffusion coefficient in the diffusive phase}

\begin{figure*}[!htb]
\centering
\includegraphics[width=9cm]{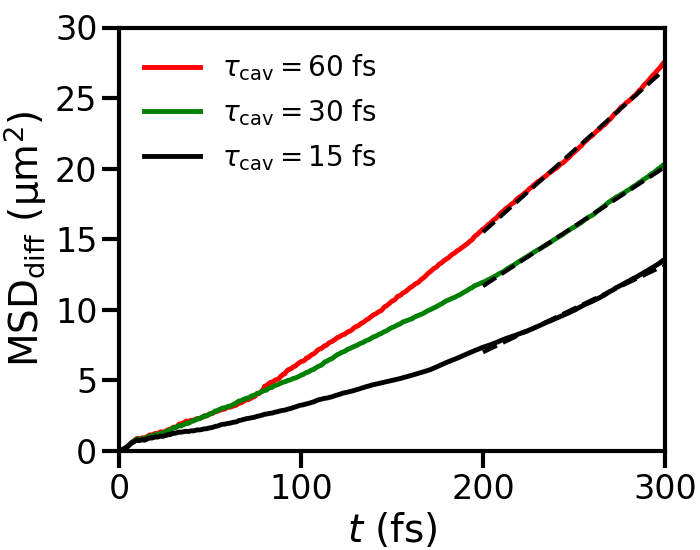}
  \caption{Mean squared displacement of the diffusive part of the total polariton wave function $|\Psi(t)|^2$ as a function of time in cavities with decay rates $\tau_{\text{cav}}$ of 60~fs~(red), 30~fs~(green), and 15~fs~(black). The dashed lines are linear fits to the last 100~fs of the MSDs.}
\label{fig:MSD_diff}
\end{figure*}

As explained in the main text, we extracted the diffusion coefficient from the wavepacket MSD. We consider only the part of the wavepacket that is moving slower than the maximum group velocity of the LP ($\upsilon_{\text{LP}}^{\text{max}}$, Figure~\ref{fig:MSD_diff}) and performed a linear fit to the last 100~fs of the trajectories:
\begin{equation}
   \text{MSD}_\text{diff}(t) = 2D\cdot t
   \label{eq:fitting_MSD}
\end{equation}
These diffusion coefficients are plotted as a function of cavity lifetime in Figure~4{\bf{c}} of the main text.

\bibliography{SI}